\documentstyle[amstex,amssymb]{article}

\catcode`\@=11
\def\numberbysection{\@addtoreset{equation}{section}
        \def\theequation{\thesection.\arabic{equation}}}
\numberbysection

\begin{document}

\newlength{\lno} \lno1.5cm \newlength{\len} \len=\textwidth%
\addtolength{\len}{-\lno}

\setcounter{page}{0}

\baselineskip7mm \renewcommand{\thefootnote}{\fnsymbol{footnote}} \newpage %
\setcounter{page}{0}

\begin{titlepage}     
\vspace{0.5cm}
\begin{center}
{\Large\bf $osp(1|2)$ off-shell Bethe ansatz equation with boundary terms}\\
\vspace{1cm}
{\large \bf  A. Lima-Santos } \\
\vspace{1cm}
{\large \em Universidade Federal de S\~ao Carlos, Departamento de F\'{\i}sica \\
Caixa Postal 676, CEP 13569-905~~S\~ao Carlos, Brasil}\\
\end{center}
\vspace{1.2cm}

\begin{abstract}
This work is concerned with the quasi-classical limit  of the boundary quantum inverse scattering method  for the
 $osp(1|2)$  vertex model with diagonal $K$-matrices.  In this limit Gaudin's Hamiltonians with boundary terms  are presented and
diagonalized. Moreover, integral representations for correlation functions are realized to be solutions of the trigonometric 
Knizhnik-Zamoldchikov equations.

\end{abstract}
\vspace{2cm}
\begin{center}
\end{center}
\vfill
\begin{center}
\small{\today}
\end{center}
\end{titlepage}

\baselineskip6mm

\newpage

\section{Introduction}

Integrable models in two-dimensional quantum field theories and the
integrable vertex models in two dimensional classical statistical mechanics
are the common link of many relevant exactly-solved quantum models in one
dimension \cite{Baxter, Gaudin, KBI, Abdalla}. Examples are the {\small XXX}%
, {\small XXZ} and the {\small XYZ} Heisenberg chains that find more and
more applications in contemporary physics. \ The quantum inverse scattering
method or algebraic Bethe ansatz is the tool to exploit this fact
systematically \cite{KBI}. The key toward this synthesis is the existence of
a central object ${\cal R}(u)$, where $u$ is a spectral parameter, acting on
the tensor product $V\otimes V$ of a given vector space $V$ and being a
solution of the celebrated quantum Yang-Baxter equation 
\begin{equation}
{\cal R}_{12}(u){\cal R}_{13}(u+v){\cal R}_{23}(v)={\cal R}_{23}(v){\cal R}%
_{13}(u+v){\cal R}_{12}(u),  \label{int.1}
\end{equation}%
in $V^{1}\otimes V^{2}\otimes V^{3}$, where ${\cal R}_{12}={\cal R}\otimes 1$%
, ${\cal R}_{23}=1\otimes {\cal R}$, etc.

The solution ${\cal R}(u)$ of (\ref{int.1}) is said to be quasi-classical if 
${\cal R}(u)$ also depends on an additional parameter $\eta $ in such a way
that%
\begin{equation}
{\cal R}(u,\eta )=1+\eta \ r(u)+{\rm o}(\eta ^{2}),  \label{int.1b}
\end{equation}%
where $1$ is the identity operator on the space $V\otimes V$. The
\textquotedblleft classical $r$-matrix\textquotedblright\ obeys the equation 
\begin{equation}
\lbrack r_{12}(u),r_{13}(u+v)+r_{23}(v)]+[r_{13}(u+v),r_{23}(v)]=0.
\label{int.2}
\end{equation}%
This equation, called the classical Yang-Baxter equation, plays an important
role in the theory of classical completely integrable systems \cite{Semenov}.

Nondegenerate solutions of (\ref{int.2}) in the tensor product of two copies
of a simple Lie algebra {\rm g} , $r_{ij}(u)\in {\rm g}_{i}\otimes {\rm g}%
_{j}$ , $i,j=1,2,3$, were classified by Belavin and Drinfeld \cite{BD}.

Gaudin models \cite{GA} constitute an important class of one dimensional
many-body systems with long-range interactions with important applications
in many branches of fields ranging from condensed matter to high energy
physics \cite{Cambia, Amico, Delft, Dukel, Seiberg}. They are related to the
classical $r$-matrices through the definition of the density Gaudin
Hamiltonians%
\begin{equation}
G_{a}=\sum_{b\neq a}^{N}r_{ab}(z_{a}-z_{b})
\end{equation}%
The commutativity condition $[G_{a},G_{b}]=0$ is a consequence of the
classical Yang-Baxter equation (\ref{int.2}).

The classical Yang-Baxter equation also has a counterpart in conformal field
theory, which can be described in the following way: in the skew-symmetric
case $r_{ji}(-u)+r_{ij}(u)=0$, it is the compatibility condition for the
system of linear differential equations 
\begin{equation}
\kappa \frac{\partial \Psi (z_{1},...,z_{N})}{\partial z_{i}}=\sum_{j\neq
i}r_{ij}(z_{i}-z_{j})\Psi (z_{1},...,z_{N})  \label{int.3}
\end{equation}%
in $N$ complex variables $z_{1},...,z_{N}$ \ for vector-valued functions $%
\Psi $ with values in the tensor space $V=V^{1}\otimes \cdots \otimes V^{N}$%
, where $\kappa $ is a coupling constant.

In the rational case \cite{BD}, very simple skew-symmetric solutions are
known: $r(u)={\rm C}_{2}/u$, where ${\rm C}_{2}\in {\rm g}\otimes {\rm g}$
is a symmetric invariant tensor of a finite dimensional Lie algebra ${\rm g}$%
. The above system of linear differential equations (\ref{int.3}) is known \
as the Knizhnik-Zamolodchikov ({\small KZ}) system of equations for the
conformal blocks of the Wess-Zumino-Novikov-Witten ({\small WZNW}) models of
the conformal field theory on the sphere \cite{KZ}.

The work of Babujian and Flume \cite{BAF} unveils one link between the
algebraic Bethe ansatz for the theory of the Gaudin models and the conformal
field theory of {\small WZWN} models. In their approach, the Bethe wave
vectors for an inhomogeneous lattice model give, in the quasi-classical
limit, solutions of the {\small KZ} equation for the case of simple Lie
algebras. For instance, in the $su(2)$ example, the quantum inverse
scattering method \cite{FT} allows one to write the following equation 
\begin{equation}
\tau (u|z)\Phi (u_{1,\cdots ,}u_{p})=\Lambda (u,u_{1},\cdots ,u_{p}|z)\Phi
(u_{1},\cdots ,u_{p})-\sum_{\alpha =1}^{p}\frac{{\cal F}_{\alpha }\Phi
^{\alpha }}{u-u_{\alpha }}.  \label{int.4}
\end{equation}%
Here $\tau (u|z)$ denotes the transfer matrix of the rational vertex model
in an inhomogeneous lattice. $\Phi ^{\alpha }$ \ meaning $\Phi ^{\alpha
}=\Phi (u_{1},\cdots u_{\alpha -1},u,u_{\alpha +1},...,u_{p})$ ; ${\cal F}%
_{\alpha }(u_{1},\cdots ,u_{p}|z)$ and $\Lambda (u,u_{1},\cdots ,u_{p}|z)$
being complex valued functions. The vanishing of the so-called unwanted
terms, ${\cal F}_{\alpha }=0$, is demanded in the usual procedure of the
algebraic Bethe ansatz \cite{Wgalleas, ALS0} in order to determine the
eigenvalues of\ $\tau (u|z)$. In this case the wave vector $\Phi
(u_{1},\cdots ,u_{p})$ becomes an eigenvector of the transfer matrix with
eigenvalue $\Lambda (u,u_{1},\cdots ,u_{p}|z)$. If we keep all unwanted
terms, i.e. ${\cal F}_{\alpha }\neq 0$, then the wave vector $\Phi $ in
general satisfies the equation (\ref{int.4}), named in \cite{B} as off-shell
Bethe ansatz equation ({\small OSBAE}).

There is a close relationship between the wave vector satisfying the {\small %
OSBAE} (\ref{int.4}) and the vector-valued solutions of the {\small KZ}
equation (\ref{int.3}): the general vector valued solution of the {\small KZ}
equation for an arbitrary simple Lie algebra was found by Schechtman and
Varchenko \cite{SV}. It can be represented as a multiple contour integral

\begin{equation}
\Psi (z_{1},\ldots ,z_{N})=\oint \cdots \oint {\cal X}(u_{1},...,u_{p}|z)%
\phi (u_{1},...,u_{p}|z)du_{1}\cdots du_{p}.  \label{int.5}
\end{equation}%
The complex variables $z_{1},...,z_{N}$ of (\ref{int.5}) are related with
the disorder parameters of the {\small OSBAE} . The vector valued function $%
\phi (u_{1},...,u_{p}|z)$ is the quasi-classical limit of the wave vector $%
\Phi (u_{1},...,u_{p}|z)$. The Bethe ansatz for the Gaudin model was derived
for any simple Lie algebra by Reshetikhin and Varchenko \cite{RV}. The
scalar function ${\cal X}(u_{1},...,u_{p}|z)$ determines the monodromy of $%
\Psi (z_{1},\ldots ,z_{N})$ and it is constructed from the quasi-classical
limit of the $\Lambda (u=z_{k};u_{1},...,u_{p}|z)$ and ${\cal F}_{\alpha
}(u_{1},\cdots ,u_{p}|z)$ functions. This representation of the $N$-point
correlation function shows a deep connection between the inhomogeneous
vertex models and the {\small WZNW \ }theory.

In \cite{BPLS} this idea has been applied on the periodic XYZ Gaudin model 
\cite{ST} in order to derive and solve the Knizhnik-Zamolodchikov-Bernard
equation for $SU(2)$-{\small WZWN}\ model on the torus. \cite{KZB}. In a
series of papers \cite{LSW, KLS1, KLS2} we have considered the $19$-vertices
models, the $osp(1|2)$ model and the twisted $A_{2}^{(2)}$ or
Izergin-Korepin model and the $sl(2|1)^{(2)}$ model, respectively. In these
works, the {\small OSBAE} procedure was used in order to find and solve the
corresponding Gaudin models, as well as, the corresponding {\small KZ}
equations. The algebraic structure of the periodic trigonometric $osp(1|2)$
Gaudin model has been studied with details in \cite{KM}.

Since the work of Sklyanin \cite{Sklyanin}, the algebraic Bethe ansatz has
been applied to various integrable models with non-trivial boundary
conditions, which are specified by $K$-matrices satisfying the reflection
equation and its dual \cite{Cherednik}. The quasi-classical expansion of the
corresponding double-row transfer matrices produces generalized Gaudin
Hamiltonians with boundaries.

Recently Gaudin models with non-trivial integrable boundaries have attracted
much interest. Initially the attention has been concentrated on Gaudin
models with diagonal $K$-matrices \cite{Hikami, Lorenzo, Gould1, Gould2}.
Nevertheless, in \cite{WenLi1} the {\small XXZ} Gaudin model was solved with
non-diagonal $K$-matrices and in \cite{WenLi2} this result was generalized
in order to solve the $A_{n-1}$ Gaudin magnets with non-diagonal $K$%
-matrices. In \ this context, in \cite{LS1} the {\small XXZ} model is also
used to solve the Gaudin magnets with non-diagonal impurities.

In this paper we present a systematic study of the algebraic structure of
the $19$-vertex model \ based on the orthosympletic Lie algebra $osp(1|2)$,
through the algebraic Bethe ansatz with diagonal $K$-matrices and we also
show how the quasi-classical limit procedure gives us the corresponding
generalized Gaudin magnets. Moreover, we will construct and solve the
trigonometric $osp(1|2)$ {\small KZ} equations.

The paper is organized as follows. In Section $2,$ we present the $osp(1|2)$
vertex model and its three diagonal $K$-matrices. In Section $3,$ the
inhomogeneous algebraic Bethe ansatz with boundary terms is presented with
details in order to derive the boundary off-shell Bethe ansatz equation for
this vertex model. In Section $4$ , the structure of the generalized Gaudin
Hamiltonians is presented. In Section $5$, we describe the corresponding
off-shell Gaudin equations taking into account the quasi-classical limit of
the results of the Section $3$. In Section $6$, data of the boundary
off-shell Gaudin equation are used to construct solutions of the
trigonometric {\small KZ} equation for the three cases. Conclusions are
reserved for Section $7$. Finally, commutation relations and some
definitions are presented in appendices.

\section{The model}

\bigskip Let $V=V_{0}\oplus V_{1}$ be a $Z_{2}$-graded vector space where $0$
and $1$ denote the even and odd parts respectively. The multiplication rules
in the graded tensor product space $V\overset{s}{\otimes }V$ differ from the
ordinary ones by the appearance of additional signs. The components of a
linear operator $A\overset{s}{\otimes }B\in V\overset{s}{\otimes }V$ result
in matrix elements of the form 
\begin{equation}
(A\overset{s}{\otimes }B)_{\alpha \beta }^{\gamma \delta }=(-1)^{p(\beta
)(p(\alpha )+p(\gamma ))}\ A_{\alpha \gamma }B_{\beta \delta }.
\label{mod.1}
\end{equation}%
The action of the graded permutation operator ${\cal P}$ on the vector $%
\left\vert \alpha \right\rangle \overset{s}{\otimes }\left\vert \beta
\right\rangle \in V\overset{s}{\otimes }V$ is defined by 
\begin{equation}
{\cal P}\ \left\vert \alpha \right\rangle \overset{s}{\otimes }\left\vert
\beta \right\rangle =(-)^{p(\alpha )p(\beta )}\left\vert \beta \right\rangle 
\overset{s}{\otimes }\left\vert \alpha \right\rangle \Longrightarrow ({\cal P%
})_{\alpha \beta }^{\gamma \delta }=(-1)^{p(\alpha )p(\beta )}\delta
_{\alpha \delta }\ \delta _{\beta \gamma }.  \label{mod.2}
\end{equation}%
The graded transposition ${\rm st}$ and the graded trace {\rm str} are
defined by 
\begin{equation}
\left( A^{{\rm st}}\right) _{\alpha \beta }=(-1)^{(p(\alpha )+1)p(\beta
)}A_{\beta \alpha },\quad {\rm str}A=\sum_{\alpha }(-1)^{p(\alpha
)}A_{\alpha \alpha }.  \label{mod.3}
\end{equation}%
where $p(\alpha )=1\ (0)$ if $\left\vert \alpha \right\rangle $ is an odd
(even) element.

For the graded case the Yang-Baxter equation 
\begin{equation}
{\cal R}_{12}(u){\cal R}_{13}(u+v){\cal R}_{23}(v)={\cal R}_{23}(v){\cal R}%
_{13}(u+v){\cal R}_{12}(u)  \label{mod.4}
\end{equation}%
and the reflection equation \cite{Sklyanin, MN}%
\begin{equation}
{\cal R}_{12}(u-v)K_{1}^{-}(u){\cal R}_{21}(u+v)K_{2}^{-}(v)=K_{2}^{-}(v)%
{\cal R}_{12}(u+v)K_{1}^{-}(u){\cal R}_{21}(u-v)  \label{mod.5}
\end{equation}%
remain the same as in the non-graded cases and we only need to change the
usual tensor product to the graded tensor product.

In general, the dual reflection equation which depends on the unitarity and
cross-unitarity relations of the ${\cal R}$-matrix takes different forms for
different models. For the model \ considered in this paper, we write the
graded dual reflection equation in the form introduced in \cite{Zhou}:%
\[
{\cal R}_{21}^{st_{1}st_{2}}(-u+v)(K_{1}^{+})^{st_{1}}(u)M_{1}^{-1}{\cal R}%
_{12}^{st_{1}st_{2}}(-u-v-2\rho )M_{1}(K_{2}^{+})^{st_{2}}(v) 
\]%
\begin{equation}
=(K_{2}^{+})^{st_{2}}(v)M_{1}{\cal R}_{12}^{st_{1}st_{2}}(-u-v-2\rho
)M_{1}^{-1}(K_{1}^{+})^{st_{1}}(u){\cal R}_{21}^{st_{1}st_{2}}(-u+v),
\label{mod.6}
\end{equation}

Now, using the relations%
\begin{equation}
{\cal R}_{12}^{st_{1}st_{2}}(u)=I_{1}R_{21}(u)I_{1},\quad {\cal R}%
_{21}^{st_{1}st_{2}}(u)=I_{1}R_{12}(u)I_{1}\quad {\rm and}\quad
IK^{+}(u)I=K^{+}(u)  \label{mod.7}
\end{equation}%
with $I={\rm diag}(1,-1,1)$ and the property $\left[ M_{1}M_{2},{\cal R}(u)%
\right] =0$ we can see that the usual isomorphism \cite{MN2}%
\begin{equation}
K^{-}(u):\rightarrow K^{+}(u)=K^{-}(-u-\rho )^{st}M.  \label{mod.8}
\end{equation}%
holds with the {\small BFB} grading. Here ${\rm st}_{i}$ denotes
super-transposition in the space $i$.

A quantum-integrable system is characterized by the monodromy matrix $T(u)$
satisfying the fundamental relation%
\begin{equation}
R(u-v)\left[ T(u)\otimes T(v)\right] =\left[ T(v)\otimes T(u)\right] R(u-v)
\label{mod.9}
\end{equation}%
where the intertwining $R$-matrix is given by $R(u)={\cal P}{\cal R}(u).$

In the framework of the quantum inverse scattering method \cite{FT}, the
simplest monodromies have become known as ${\cal L}$ operators, the Lax
operators, here defined by ${\cal L}_{aq}(u)={\cal R}_{aq}(u)$, where the
subscript $a$ represents the auxiliary (horizontal) space, and $q$
represents the quantum (vertical) space. The monodromy matrix $T(u)$ is
defined as the matrix product of $N$ \ Lax operators on all sites of the
lattice,%
\begin{equation}
T(u)={\cal L}_{aN}(u){\cal L}_{aN-1}(u)\cdots {\cal L}_{a1}(u).
\label{mod.10}
\end{equation}

The main result for open boundaries integrability is: if the boundary
equations are satisfied, then the Sklyanin's transfer matrix \cite{Sklyanin}%
\begin{equation}
t(u)={\rm str}_{a}\left( K^{+}(u)T(u)K^{-}(u)T^{-1}(-u)\right) ,
\label{mod.11}
\end{equation}%
forms a commuting collection of operators in the quantum space%
\begin{equation}
\left[ t(u),t(v)\right] =0,\qquad \forall u,v  \label{mod.12}
\end{equation}

The commutativity of $t(u)$ can be proved by using the unitarity and
crossing-unitarity relations, the reflection equation and the dual
reflection equation. In particular, it implies the integrability of an open
quantum spin chain whose Hamiltonian (with $K^{-}(0)=1$) is given by \cite%
{Sklyanin}%
\begin{equation}
H=\sum_{k=1}^{N-1}H_{k,k+1}+\frac{1}{2}\left. \frac{dK_{1}^{-}(u)}{du}%
\right\vert _{u=0}+\frac{{\rm str}_{0}K_{0}^{+}(0)H_{N,0}}{{\rm str}K^{+}(0)}%
,  \label{mod.13}
\end{equation}%
where the two-site terms are given by%
\begin{equation}
H_{k,k+1}=\left. \frac{d}{du}R_{k,k+1}(u)\right\vert _{u=0}  \label{mod.14}
\end{equation}%
in the standard fashion.

The trigonometric solution of the graded Yang-Baxter equation (\ref{mod.4})
corresponding to $osp(1|2)$ \ in the fundamental representation has the form%
\begin{equation}
{\cal R}(u)=\frac{1}{x_{2}}\left( 
\begin{array}{ccccccccc}
x_{1} & 0 & 0 & 0 & 0 & 0 & 0 & 0 & 0 \\ 
0 & x_{2} & 0 & x_{5} & 0 & 0 & 0 & 0 & 0 \\ 
0 & 0 & x_{3} & 0 & x_{6} & 0 & x_{7} & 0 & 0 \\ 
0 & y_{5} & 0 & x_{2} & 0 & 0 & 0 & 0 & 0 \\ 
0 & 0 & y_{6} & 0 & x_{4} & 0 & x_{6} & 0 & 0 \\ 
0 & 0 & 0 & 0 & 0 & x_{2} & 0 & x_{5} & 0 \\ 
0 & 0 & y_{7} & 0 & y_{6} & 0 & x_{3} & 0 & 0 \\ 
0 & 0 & 0 & 0 & 0 & y_{5} & 0 & x_{2} & 0 \\ 
0 & 0 & 0 & 0 & 0 & 0 & 0 & 0 & x_{1}%
\end{array}%
\right)  \label{mod.15}
\end{equation}%
with non-zero entries \cite{BS}:%
\begin{eqnarray}
x_{1}(u) &=&\sinh (u+2\eta )\sinh (u+3\eta ),\quad x_{2}(u)=\sinh u\sinh
(u+3\eta )  \nonumber \\
x_{3}(u) &=&\sinh u\sinh (u+\eta ),\quad x_{4}(u)=\sinh u\sinh (u+3\eta
)-\sinh 2\eta \sinh 3\eta  \nonumber \\
x_{5}(u) &=&{\rm e}^{-u}\sinh 2\eta \sinh (u+3\eta ),\quad y_{5}(u)={\rm e}%
^{u}\sinh 2\eta \sinh (u+3\eta )  \nonumber \\
x_{6}(u) &=&-\epsilon {\rm e}^{-u-2\eta }\sinh 2\eta \sinh u,\quad
y_{6}(u)=\epsilon {\rm e}^{u+2\eta }\sinh 2\eta \sinh u  \nonumber \\
x_{7}(u) &=&{\rm e}^{-u}\sinh 2\eta \left( \sinh (u+3\eta )+{\rm e}^{-\eta
}\sinh u\right)  \nonumber \\
y_{7}(u) &=&{\rm e}^{u}\sinh 2\eta \left( \sinh (u+3\eta )+{\rm e}^{\eta
}\sinh u\right)  \label{mod.16}
\end{eqnarray}%
satisfying the properties%
\begin{eqnarray}
{\rm regularity} &:&{\cal R}_{12}(0)=f(0)^{1/2}P_{12},  \nonumber \\
{\rm unitarity} &:&{\cal R}_{12}(u){\cal R}_{12}^{st_{1}st_{2}}(-u)=f(u), 
\nonumber \\
{\rm PT-symmetry} &:&P_{12}{\cal R}_{12}(u)P_{12}={\cal R}%
_{12}^{st_{1}st_{2}}(u),  \nonumber \\
{\rm cros}\text{sin}{\rm g-symmetry} &:&{\cal R}_{12}(u)=U_{1}{\cal R}%
_{12}^{st_{2}}(-u-\rho )U_{1}^{-1},  \label{mod.17}
\end{eqnarray}%
where $\epsilon =\pm 1,\ $ $f(u)=\frac{x_{1}(u)}{x_{2}(u)}\frac{x_{1}(-u)}{%
x_{2}(-u)}$ and $\rho $ is the crossing parameter and $U$ determines the
crossing matrix $M=U^{t}U=M^{t}.$ Here we have assumed that the grading of
threefold space is $p(1)=p(3)=0$ and $p(2)=1$, the {\small BFB} grading and
we will choose the solution with $\epsilon =1$. Moreover, we have verified
that the three different sets of equations associated with the gradings 
{\small BFB}, {\small FBB} and {\small BBF} are mutually equivalent.

This ${\cal R}$-matrix is regular and unitary. It is $PT$-symmetric and
crossing-symmetric, with\ $\rho =3\eta $ and%
\begin{equation}
M=\left( 
\begin{array}{ccc}
{\rm e}^{-2\eta } & 0 & 0 \\ 
0 & 1 & 0 \\ 
0 & 0 & {\rm e}^{2\eta }%
\end{array}%
\right) .  \label{mod.18}
\end{equation}

Diagonal solutions $K^{-}(u)$ \ for the reflection equation (\ref{mod.5})
have been obtained in \cite{LS2}. It turns out that there are three
solutions without free parameters, being $K^{-}(u)=1$, $K^{-}(u)=F^{+}$ and $%
K^{-}(u)=F^{-}$, with%
\begin{equation}
F^{\pm }=\left( 
\begin{array}{ccc}
\mp {\rm e}^{-2u}f^{(\pm )}(u) & 0 & 0 \\ 
0 & 1 & 0 \\ 
0 & 0 & \mp {\rm e}^{2u}f^{(\pm )}(u)%
\end{array}%
\right) ,  \label{mod.19}
\end{equation}%
where we have defined%
\begin{equation}
f^{(+)}(u)=\frac{\sinh (u+\frac{3}{2}\eta )}{\sinh (u-\frac{3}{2}\eta )}%
,\quad f^{(-)}(u)=\frac{\cosh (u+\frac{3}{2}\eta )}{\cosh (u-\frac{3}{2}\eta
)}.  \label{mod.20}
\end{equation}%
By the automorphism (\ref{mod.8}), three solutions $K^{+}(u)$ follow as $%
K^{+}(u)=M$, $K^{+}(u)=G^{+}$ and $K^{+}(u)=G^{-}$, with%
\begin{equation}
G^{\pm }=\left( 
\begin{array}{ccc}
\mp {\rm e}^{2u+4\eta }g^{(\pm )}(u) & 0 & 0 \\ 
0 & 1 & 0 \\ 
0 & 0 & \mp {\rm e}^{-2u-4\eta }g^{(\pm )}(u)%
\end{array}%
\right) ,  \label{mod.21}
\end{equation}%
where we have defined%
\begin{equation}
g^{(+)}(u)=\frac{\sinh (u+\frac{3}{2}\eta )}{\sinh (u+\frac{9}{2}\eta )}%
,\quad g^{(-)}(u)=\frac{\cosh (u+\frac{3}{2}\eta )}{\cosh (u+\frac{9}{2}\eta
)}.  \label{mod.22}
\end{equation}

We have thus nine possibilities for the commuting transfer matrix (\ref%
{mod.11}). We will only consider three types of boundary solutions, one for
each pair ($K^{-}(u),K^{+}(u)$) defined by the automorphism (\ref{mod.8}): $%
(1,M)$, $(F^{+},G^{+})$ and $(F^{-},G^{-})$.

\section{The boundary off-shell Bethe equation}

The boundary algebraic Bethe ansatz for a $N$-site inhomogeneous double-row
transfer matrix is obtained from the homogeneous ones by a local shifting of
the spectral parameter $u$ in the monodromy matrix $T(u)$ and in its
reflection $T^{-1}(-u)$:%
\begin{equation}
T_{ia}(u|z)=\sum_{k_{1},...,k_{N-1}=1}^{3}{\cal L}_{ik_{1}}^{(N)}(u-z_{N},%
\eta )\overset{s}{\otimes }{\cal L}_{k_{1}k_{2}}^{(N-1)}(u-z_{N-1},\eta )%
\overset{s}{\otimes }\cdots \overset{s}{\otimes }{\cal L}%
_{k_{N-1}a}^{(1)}(u-z_{1},\eta )  \label{baba.1}
\end{equation}%
and%
\begin{equation}
T_{bj}^{-1}(-u|z)=\sum_{k_{1},...,k_{N-1}=1}^{3}{\cal L}_{bk_{1}}^{\prime
(1)}(u+z_{1},\eta )\overset{s}{\otimes }{\cal L}_{k_{1}k_{2}}^{\prime
(2)}(u+z_{2},\eta )\overset{s}{\otimes }\cdots \overset{s}{\otimes }{\cal L}%
_{k_{N-1}j}^{\prime (N)}(u+z_{N},\eta ).  \label{baba.2}
\end{equation}%
where 
\begin{equation}
{\cal L}_{ij}^{\prime (n)}(u+z_{n},\eta )=\frac{1}{f(u+z_{n})}{\cal L}%
_{ij}^{(n)}(-u-z_{n},-\eta ),  \label{baba.3}
\end{equation}%
are the $3\times 3$ matrices (obtained from the identification ${\cal L}%
_{aq}(u)={\cal R}_{aq}(u)$\ ) acting on the $n${\rm th} site of the lattice.

\bigskip The double-row monodromy matrix $U(u)$ can be written as a $3\times
3$ matrix

\begin{equation}
U(u|z)=T(u|z)K^{-}(u)T^{-1}(-u|z)=\left( 
\begin{array}{ccc}
U_{11}(u|z) & U_{12}(u|z) & U_{13}(u|z) \\ 
U_{21}(u|z) & U_{22}(u|z) & U_{23}(u|z) \\ 
U_{31}(u|z) & U_{32}(u|z) & U_{33}(u|z)%
\end{array}%
\right)  \label{baba.4}
\end{equation}%
and it is more convenient to introduce new operators:

\begin{eqnarray}
{\cal D}_{1}(u|z) &=&U_{11}(u|z),  \nonumber \\
{\cal D}_{2}(u|z) &=&U_{22}(u|z)-f_{1}(u){\cal D}_{1}(u|z),  \nonumber \\
{\cal D}_{3}(u|z) &=&U_{33}(u|z)-f_{2}(u){\cal D}_{1}(u|z)-f_{3}(u){\cal D}%
_{2}(u|z),  \nonumber \\
{\cal B}_{i}(u|z) &=&U_{ij}(u|z)\ (i<j),\qquad {\cal C}_{i}(u|z)=U_{ij}(u|z)%
\ (i>j),\qquad i,j=1,2,3  \label{baba.5}
\end{eqnarray}

The usual reference state is 
\begin{equation}
\left\vert 0\right\rangle =\left( 
\begin{array}{c}
1 \\ 
0 \\ 
0%
\end{array}%
\right) \overset{s}{\otimes }\left( 
\begin{array}{c}
1 \\ 
0 \\ 
0%
\end{array}%
\right) \overset{s}{\otimes }\cdots \overset{s}{\otimes }\left( 
\begin{array}{c}
1 \\ 
0 \\ 
0%
\end{array}%
\right)  \label{baba.7}
\end{equation}%
which is a highest vector of ${\cal U}(u)$%
\begin{equation}
{\cal U}(u)\left\vert 0\right\rangle =\left( 
\begin{array}{ccc}
{\cal X}_{1}(u|z)\left\vert 0\right\rangle & \ast & \ast \\ 
0 & {\cal X}_{2}(u|z)\left\vert 0\right\rangle & \ast \\ 
0 & 0 & {\cal X}_{3}(u|z)\left\vert 0\right\rangle%
\end{array}%
\right)  \label{baba.8}
\end{equation}%
where the symbol $(\ast )$ denotes states different from $0$ and $\left\vert
0\right\rangle $ and%
\begin{eqnarray}
{\cal X}_{1}(u|z) &=&k_{11}^{-}(u)\prod\limits_{a=1}^{N}\frac{x_{1}(u-z_{a})%
}{x_{2}(u-z_{a})}\frac{x_{1}(u+z_{a})}{x_{2}(u+z_{a})}\frac{1}{f(u+z_{a})}, 
\nonumber \\
{\cal X}_{2}(u|z) &=&\left[ k_{22}^{-}(u)-k_{11}^{-}(u)f_{1}(u)\right]
\prod\limits_{a=1}^{N}\frac{1}{f(u+z_{a})},  \nonumber \\
{\cal X}_{3}(u|z) &=&\left[
k_{33}^{-}(u)-k_{22}^{-}(u)f_{3}(u)-k_{11}^{-}(u)f_{4}(u)\right]
\prod\limits_{a=1}^{N}\frac{x_{3}(u-z_{a})}{x_{2}(u-z_{a})}\frac{%
x_{3}(u+z_{a})}{x_{2}(u+z_{a})}\frac{1}{f(u+z_{a})}.  \label{baba.9}
\end{eqnarray}%
where%
\begin{eqnarray}
f_{1}(u) &=&\frac{y_{5}(2u)}{x_{1}(2u)},\qquad f_{3}(u)=-\frac{%
x_{1}(2u)y_{5}(2u)-x_{5}(2u)y_{7}(2u)}{x_{1}(2u)x_{4}(2u)+x_{5}(2u)y_{5}(2u)}%
,  \nonumber \\
f_{2}(u) &=&\frac{y_{7}(2u)}{x_{1}(2u)},\qquad f_{4}(u)=\frac{%
x_{4}(2u)y_{7}(2u)+y_{5}^{2}(2u)}{x_{1}(2u)x_{4}(2u)+x_{5}(2u)y_{5}(2u)}.
\label{baba.6}
\end{eqnarray}

Using the $K^{+}(u)$ reflection matrices (\ref{mod.21}) and the {\small BFB}
grading, the double-row transfer matrix (\ref{mod.11}) has the form%
\begin{eqnarray}
t(u|z)
&=&k_{11}^{+}(u)U_{11}(u|z)-k_{22}^{+}(u)U_{22}(u|z)+k_{33}^{+}(u)U_{33}(u|z)
\nonumber \\
&=&\Omega _{1}(u){\cal D}_{1}(u|z)+\Omega _{2}(u){\cal D}_{2}(u|z)+\Omega
_{3}(u){\cal D}_{3}(u|z)  \label{baba.10}
\end{eqnarray}%
where%
\begin{eqnarray}
\Omega _{1}(u) &=&k_{11}^{+}(u)-f_{1}(u)k_{22}^{+}(u)+f_{2}(u)k_{33}^{+}(u),
\nonumber \\
\Omega _{2}(u) &=&-k_{22}^{+}(u)+f_{3}(u)k_{33}^{+}(u),  \nonumber \\
\Omega _{3}(u) &=&k_{33}^{+}(u).  \label{baba.11}
\end{eqnarray}%
The \ monodromy matrix ${\cal U}(u)$ satisfies the relation%
\begin{equation}
{\cal R}_{12}(u-v){\cal U}_{1}(u){\cal R}_{21}(u+v){\cal U}_{2}(v)={\cal U}%
_{2}(v){\cal R}_{12}(u+v){\cal U}_{1}(u){\cal R}_{21}(u-v)  \label{baba.11a}
\end{equation}%
from which we obtain the commutation relations for its entries. The
commutation relations which participate effectively in the Bethe ansatz
construction are presented in the appendix A.

The $n$-particle state is defined as an operator valued function through the
recurrence relation

\begin{eqnarray}
\Psi _{n}(u,\ldots ,u_{n}|z) &=&{\cal B}_{1}(u_{1}|z)\Psi
_{n-1}(u_{2},\ldots ,u_{n}|z)  \nonumber \\
&&+{\cal B}_{2}(u_{1}|z)\sum_{i=2}^{n}{\cal X}_{1}(u_{i}|z)\digamma
_{1}^{(i)}(u_{1},\ldots ,u_{n})\Psi _{n-2}(u_{2},\ldots ,\overset{\vee }{%
u_{i}},\ldots ,u_{n}|z)  \nonumber \\
&&+{\cal B}_{2}(u_{1}|z)\sum_{i=2}^{n}{\cal X}_{2}(u_{i}|z)\digamma
_{2}^{(i)}(u_{1},\ldots ,u_{n})\Psi _{n-2}(u_{2},\ldots ,\overset{\vee }{%
u_{i}},\ldots ,u_{n}|z)  \label{baba.12}
\end{eqnarray}%
with $\Psi _{0}=\left\vert 0\right\rangle $ and $\Psi _{1}(u_{1})={\cal B}%
(u_{1}|z)\left\vert 0\right\rangle $. Here $\overset{\vee }{u_{i}}$ denotes
that the rapidity $u_{i}$ is absent.

It was shown in \cite{China} that this operator is normal ordered satisfying 
$n-1$ exchange conditions%
\begin{equation}
\Psi _{n}(u_{1},\ldots ,u_{i},u_{i+1},\ldots ,u_{n}|z)=\omega
(u_{i},u_{i+1})\Psi _{n}(u_{1},\ldots ,u_{i+1},u_{i},\ldots ,u_{n}|z)
\label{baba.13}
\end{equation}%
provided the functions $\digamma _{\alpha }^{(i)}(u_{1},\ldots ,u_{n})$ are
given by 
\begin{equation}
\digamma _{\alpha }^{(i)}(u_{1},\ldots
,u_{n})=\prod\limits_{j=2}^{i-1}\omega (u_{j},u_{i})\prod_{k=2,k\neq
i}^{n}a_{\alpha 1}(u_{i},u_{k})G_{d_{\alpha }}(u_{1},u_{i}),\qquad (\alpha
=1,2)  \label{baba.14}
\end{equation}%
where%
\begin{eqnarray}
&&\omega (u,v)=-\frac{x_{3}(u-v)x_{4}(u-v)-x_{6}(u-v)y_{6}(u-v)}{%
x_{1}(u-v)x_{3}(u-v)}  \nonumber \\
&&\omega (u_{2},v)\omega (u,v)=1
\end{eqnarray}%
and 
\begin{equation}
G_{d_{1}}(u,v)=\frac{x_{6}(u-v)}{x_{3}(u-v)}\frac{x_{2}(2v)}{x_{1}(2v)}%
,\quad G_{d_{2}}(u,v)=-\frac{x_{6}(u+v)}{x_{2}(u+v)}  \label{baba.15b}
\end{equation}

Using the commutation relations listed in the appendix A, one can find the
action of the operators ${\cal D}_{\alpha }(u|z)$ , $a=1,2,3$ , on the $n$%
-particle state. In order to do so we can start with the one-particle state,
then the two-particle state and so on. This procedure is very extensive and
tedious but the final result can be presented in a compact form%
\begin{eqnarray}
{\cal D}_{\alpha }(u|z)\Psi _{n}(u_{1},\ldots ,u_{n}|z) &=&{\cal X}_{\alpha
}(u|z)\prod\limits_{i=1}^{n}a_{\alpha 1}(u,u_{i})\Psi _{n}(u_{1},\ldots
,u_{n}|z)  \nonumber \\
&&+\sum_{i=1}^{n}\prod\limits_{j=1}^{i-1}\omega (u_{j},u_{i})\left( {\cal X}%
_{1}(u_{i}|z)a_{\alpha 2}(u,u_{i})\prod\limits_{k\neq
i}^{n}a_{11}(u_{i},u_{k})\right.  \nonumber \\
&&+\left. {\cal X}_{2}(u_{i}|z)a_{\alpha 3}(u,u_{i})\prod\limits_{k\neq
i}^{n}a_{21}(u_{i},u_{k})\right) {\cal B}_{1}(u|z)\Psi _{n-1}(\overset{\vee }%
{u_{i}}|z)  \nonumber \\
&&+(1-\delta _{\alpha ,1})\sum_{i=1}^{n}\prod\limits_{j=1}^{i-1}\omega
(u_{j},u_{i})\left( {\cal X}_{1}(u_{i}|z)a_{\alpha
4}(u,u_{i})\prod\limits_{k\neq i}^{n}a_{11}(u_{i},u_{k})\!\right.  \nonumber
\\
&&+\left. {\cal X}_{2}(u_{i}|z)a_{\alpha 5}(u,u_{i})\prod\limits_{k\neq
i}^{n}a_{21}(u_{i},u_{k})\right) {\cal B}_{3}(u|z)\Psi _{n-1}(\overset{\vee }%
{u_{i}}|z)  \nonumber \\
&&+\sum_{i=1}^{n-1}\sum_{j=i+1}^{n}\left\{ {\cal X}_{1}(u_{i}|z){\cal X}%
_{1}(u_{j}|z)\prod_{k\neq i,j}^{n}a_{11}(u_{i},u_{k})\prod_{l\neq
i,j}^{n}a_{11}(u_{j},u_{l})H_{\alpha 1}(u_{i},u_{j})\right.  \nonumber \\
&&+{\cal X}_{2}(u_{i}|z){\cal X}_{1}(u_{j}|z)\prod_{k\neq
i,j}^{n}a_{21}(u_{i},u_{k})\prod_{l\neq i,j}^{n}a_{11}(u_{j},u_{l})H_{\alpha
2}(u_{i},u_{j})  \nonumber \\
&&+{\cal X}_{1}(u_{i}|z){\cal X}_{2}(u_{j}|z)\prod_{k\neq
i,j}^{n}a_{11}(u_{i},u_{k})\prod_{l\neq i,j}^{n}a_{21}(u_{j},u_{l})H_{\alpha
3}(u_{i},u_{j})  \nonumber \\
&&+\left. {\cal X}_{2}(u_{i}|z){\cal X}_{2}(u_{j}|z)\prod_{k\neq
i,j}^{n}a_{21}(u_{i},u_{k})\prod_{l\neq i,j}^{n}a_{21}(u_{j},u_{l})H_{\alpha
4}(u_{i},u_{j})\right\}  \nonumber \\
&&\times \prod\limits_{k=1}^{i-1}\omega (u_{k},u_{i})\prod\limits_{l=1\neq
i}^{j-1}\omega (u_{l},u_{j}){\cal B}_{2}(u|z)\Psi _{n-2}(\overset{\vee }{%
u_{i}},\overset{\vee }{u_{j}}|z)  \label{baba.16}
\end{eqnarray}

The action of the transfer matrix $t(u|z)$ on this $n$-particle state
defines the so-called off-shell Bethe ansatz equation \cite{B}. In our case
it has the following form 
\begin{equation}
t(u|z)\Psi _{n}(u_{1},\ldots ,u_{n}|z)=\Lambda _{n}\Psi _{n}(u_{1},\ldots
,u_{n}|z)+\sum_{j=1}^{n}{\cal F}_{j}^{(n-1)}\Psi
_{(n-1)}^{j}+\sum_{j=2}^{n}\sum_{l=1}^{j-1}{\cal F}_{lj}^{(n-2)}\Psi
_{(n-2)}^{lj}  \label{baba.17}
\end{equation}%
Let us explain a bit more the right hand side terms of this equation: In the
first term the Bethe vectors (\ref{baba.12}) are multiplied by $c$-numbers $%
\Lambda _{n}=\Lambda _{n}(u,u_{1},...,u_{n}|z)$ given by%
\begin{equation}
\Lambda _{n}=\sum_{\alpha =1}^{3}\Omega _{\alpha }(u){\cal X}_{\alpha
}(u|z)\prod\limits_{k=1}^{n}a_{\alpha 1}(u,u_{k}).  \label{baba.18}
\end{equation}%
The second term is a sum of new operator valued functions%
\begin{equation}
\Psi _{(n-1)}^{j}=\prod\limits_{k=1}^{j-1}\omega (u_{k},u_{j})\left(
\sum_{\alpha =1}^{3}\Omega _{\alpha }(u)a_{\alpha 2}(u,u_{j}){\cal B}%
_{1}(u|z)+\sum_{\alpha =2}^{3}\Omega _{\alpha }(u)a_{\alpha 4}(u,u_{j}){\cal %
B}_{3}(u|z)\right) \Psi _{(n-1)}(\overset{\vee }{u_{j}})  \label{baba.19}
\end{equation}%
multiplied by scalar functions 
\begin{equation}
{\cal F}_{j}^{(n-1)}={\cal X}_{1}(u_{j}|z)\prod\limits_{k\neq
j}^{n}a_{11}(u_{j},u_{k})+\Theta (u_{j}){\cal X}_{2}(u_{j}|z)\prod\limits_{k%
\neq j}^{n}a_{21}(u_{j},u_{k})  \label{baba.20}
\end{equation}%
where we have defined%
\begin{equation}
\Theta (u_{j})=\frac{%
\displaystyle \sum_{\alpha=1}^{3}%
\Omega _{\alpha }(u)a_{\alpha 3}(u,u_{j})}{%
\displaystyle \sum_{\alpha=1}^{3}%
\Omega _{\alpha }(u)a_{\alpha 2}(u,u_{j})}=\frac{%
\displaystyle \sum_{\alpha=2}^{3}%
\Omega _{\alpha }(u)a_{\alpha 5}(u,u_{j})}{%
\displaystyle \sum_{\alpha=2}^{3}%
\Omega _{\alpha }(u)a_{\alpha 4}(u,u_{j})}.  \label{baba.21}
\end{equation}%
Finally, the last term is a coupled sum of a third type of operator valued
functions%
\begin{equation}
\Psi _{(n-2)}^{lj}=\prod\limits_{k=1}^{l-1}\omega
(u_{k},u_{l})\prod\limits_{k=1,\neq l}^{j-1}\omega (u_{k},u_{j}){\cal B}%
_{2}(u|z)\Psi _{(n-2)}(\overset{\vee }{u_{l}},\overset{\vee }{u_{j}})
\label{baba.22}
\end{equation}%
with intricate coefficients 
\begin{eqnarray}
{\cal F}_{lj}^{(n-2)} &=&{\cal X}_{1}(u_{l}|z){\cal X}_{1}(u_{j}|z)\prod_{k%
\neq l}^{n}a_{11}(u_{l},u_{k})\prod_{k\neq
,j}^{n}a_{11}(u_{j},u_{k})\sum_{\alpha =1}^{3}\Omega _{\alpha }(u)H_{\alpha
1}(u_{l},u_{j})  \nonumber \\
&&+{\cal X}_{2}(u_{l}|z){\cal X}_{1}(u_{j}|z)\prod_{k\neq
l}^{n}a_{21}(u_{l},u_{k})\prod_{k\neq j}^{n}a_{11}(u_{j},u_{k})\sum_{\alpha
=1}^{3}\Omega _{\alpha }(u)H_{\alpha 2}(u_{l},u_{j})  \nonumber \\
&&+{\cal X}_{1}(u_{l}|z){\cal X}_{2}(u_{j}|z)\prod_{k\neq
l}^{n}a_{11}(u_{l},u_{k})\prod_{k\neq j}^{n}a_{21}(u_{j},u_{k})\sum_{\alpha
=1}^{3}\Omega _{\alpha }(u)H_{\alpha 3}(u_{l},u_{j})  \nonumber \\
&&+{\cal X}_{2}(u_{l}|z){\cal X}_{2}(u_{j}|z)\prod_{k\neq
l}^{n}a_{21}(u_{l},u_{k})\prod_{k\neq j}^{n}a_{21}(u_{j},u_{k})\sum_{\alpha
=1}^{3}\Omega _{\alpha }(u)H_{\alpha 4}(u_{l},u_{j})  \label{baba.23}
\end{eqnarray}

The functions $a_{ij}(u,v)$ are amplitudes which came from the commutation
relations (appendix A), while $H_{\alpha j}(u_{p},u_{q})$ \ ($\alpha
=1,2,3,\ j=1,2,3,4$) are cumbersome functions of this amplitudes which we
have left to the appendix B.

In the usual algebraic Bethe ansatz method, the next step consist in impose
the vanishing of these unwanted terms in order to get the eigenvalue problem
for the inhomogeneous transfer matrix $t(u|z)$: We impose ${\cal F}%
_{i}^{(n-1)}=0$, which implies ${\cal F}_{lj}^{(n-2)}=0$. From these
elimination conditions we can see that $\Psi _{n}(u_{1},\ldots ,u_{n}|z)$ is
an eigenstate of $t(u|z)$ with eigenvalue (\ref{baba.18}) provided that the
rapidities $u_{k}$ are solutions of the inhomogeneous Bethe equations 
\begin{equation}
\frac{{\cal X}_{1}(u_{k}|z)}{{\cal X}_{2}(u_{k}|z)}=-\Theta
(u_{k})\prod\limits_{j=1,j\neq k}^{n}\frac{a_{21}(u_{k},u_{j})}{%
a_{11}(u_{k},u_{j})},\qquad (k=1,2,\ldots ,n)  \label{baba.24}
\end{equation}%
where the factors $\Theta (u_{k})$ \ are easily identified with the phase
shifts at the boundaries.

\section{The Gaudin Hamiltonians}

We recall that the $osp(1|2)$ has three even (bosonic) generators $H,$\ $%
X^{\pm }$ and two odd (fermionic) generators $V^{\pm }$ , whose
non-vanishing commutation relations in the Cartan-Weyl basis reads as 
\begin{eqnarray}
\lbrack H,X^{\pm }] &=&\pm X^{\pm },\quad \lbrack X^{+},X^{-}]=2H,  \nonumber
\\
\lbrack H,V^{\pm }] &=&\pm \frac{1}{2}V^{\pm },\quad \lbrack X^{\pm },V^{\mp
}]=V^{\pm },\quad \lbrack X^{\pm },V^{\pm }]=0,  \nonumber \\
\{V^{\pm },V^{\pm }\} &=&\pm \frac{1}{2}X^{\pm },\quad \{V^{+},V^{-}\}=-%
\frac{1}{2}H.  \label{str.1}
\end{eqnarray}%
The quadratic Casimir operator is 
\begin{equation}
C_{2}=H^{2}+\frac{1}{2}\{X^{+},X^{-}\}+[V^{+},V^{-}],  \label{str.2}
\end{equation}%
where $\{\cdot \ ,\cdot \}$ denotes the anticommutator and $[\cdot \ ,\cdot
] $ the commutator.

The irreducible finite-dimensional representations $\rho _{j}$ with the
highest weight vector are parametrized by half-integer $s=j/2$ or by the
integer $j\in N$. Their dimension is ${\rm dim}(\rho _{j})=2j+1$ and the
corresponding eigenvalue of the quadratic Casimir is $j(j+1)/4=s(s+1/2)$.

The fundamental representation has $s=1/2$ and its matrix realization in the 
{\small BFB} grading is 
\begin{eqnarray}
H &=&\frac{1}{2}\left( 
\begin{array}{lll}
1 & 0 & \ \ 0 \\ 
0 & 0 & \ \ 0 \\ 
0 & 0 & -{}1%
\end{array}%
\right) ,\ \ X^{+}=\left( 
\begin{array}{lll}
0 & 0 & 1 \\ 
0 & 0 & 0 \\ 
0 & 0 & 0%
\end{array}%
\right) ,\ X^{-}=\left( 
\begin{array}{lll}
0 & 0 & 0 \\ 
0 & 0 & 0 \\ 
1 & 0 & 0%
\end{array}%
\right) ,  \nonumber \\
V^{+} &=&\frac{1}{2}\left( 
\begin{array}{lll}
0 & 1 & 0 \\ 
0 & 0 & 1 \\ 
0 & 0 & 0%
\end{array}%
\right) ,\ V^{-}=\frac{1}{2}\left( 
\begin{array}{lll}
\ \ 0 & 0 & 0 \\ 
-1 & 0 & 0 \\ 
\ \ 0 & 1 & 0%
\end{array}%
\right) .  \label{str.4}
\end{eqnarray}

In this section we will consider the theory of the Gaudin model with
boundary terms. To do this we need to calculate the quasi-classical limit of
the results presented in the previous section.

We can begin expanding (up to an appropriate order in $\eta $), the \
diagonal Lax operator entries of the monodromy matrix $T(u|z)$%
\begin{eqnarray}
{\cal L}_{11}^{(n)} &=&{\cal I}_{n}+2\eta \ \coth (u-z_{n}){\cal H}%
_{n}+2\eta ^{2}\left( {\cal H}_{n}^{2}+\frac{3}{2}\frac{{\cal H}_{n}^{2}-%
{\cal H}_{n}}{\sinh (u-z_{n})^{2}}\right) +{\rm o}(\eta ^{3})  \nonumber \\
&&  \nonumber \\
{\cal L}_{22}^{(n)} &=&{\cal I}_{n}-2\eta ^{2}\frac{3{\cal I}_{n}-3{\cal H}%
_{n}^{2}}{\sinh (u-z_{n})^{2}}+{\rm o}(\eta ^{3})  \nonumber \\
&&  \nonumber \\
{\cal L}_{33}^{(n)} &=&{\cal I}_{n}-2\eta \ \coth (u-z_{n}){\cal H}%
_{n}+2\eta ^{2}\left( {\cal H}_{n}^{2}+\frac{3}{2}\frac{{\cal H}_{n}^{2}+%
{\cal H}_{n}}{\sinh (u-z_{n})^{2}}\right) +{\rm o}(\eta ^{3}).  \label{str.5}
\end{eqnarray}%
For the elements out of the diagonal we have considered a different order in 
$\eta $ 
\begin{eqnarray}
{\cal L}_{12}^{(n)} &=&-2\eta \ \frac{{\rm e}^{-u+z_{n}}}{\sinh (u-z_{n})}%
{\cal V}_{n}^{-}+{\rm o}(\eta ^{2}),\quad \ {\cal L}_{21}^{(n)}=2\eta \ 
\frac{{\rm e}^{u-z_{n}}}{\sinh (u-z_{n})}{\cal V}_{n}^{+}+{\rm o}(\eta ^{2}),
\nonumber \\
&&  \nonumber \\
{\cal L}_{23}^{(n)} &=&2\eta \ \frac{{\rm e}^{-u+z_{n}}}{\sinh (u-z_{n})}%
{\cal V}_{n}^{-}+{\rm o}(\eta ^{2}),\quad \quad {\cal L}_{32}^{(n)}=2\eta \ 
\frac{{\rm e}^{u-z_{n}}}{\sinh (u-z_{n})}{\cal V}_{n}^{+}+{\rm o}(\eta ^{2}),
\nonumber \\
&&  \nonumber \\
{\cal L}_{13}^{(n)} &=&2\eta \ \frac{{\rm e}^{-u+z_{n}}}{\sinh (u-z_{n})}%
{\cal X}_{n}^{-}+{\rm o}(\eta ^{2}),\quad \quad {\cal L}_{31}^{(n)}=2\eta \ 
\frac{{\rm e}^{u-z_{n}}}{\sinh (u-z_{n})}{\cal X}_{n}^{+}+{\rm o}(\eta ^{2}).
\label{str.6}
\end{eqnarray}%
where ${\cal V}^{\pm }=2V^{\pm }$, ${\cal X}^{\pm }=2X^{\pm }$ and ${\cal H}%
=2H$.

The corresponding expressions for the reflected monodromy matrix $%
T^{-1}(-u|z)$ can be read from (\ref{str.5}) and (\ref{str.6}) by the
substitution $\eta \rightarrow -\eta $ and $u\rightarrow -u.$ For sake of
simplicity we can remove the global factor by considering its expansion as $%
\prod (1/f)\approx 1+{\rm o}(\eta ^{2})$ instead of a normalization. \ 

The quasi-classical expansions for the entries of the $K$ matrices (\ref%
{mod.18})-(\ref{mod.22}) will be written in the form%
\begin{equation}
k_{\alpha \alpha }^{\pm }(u)=k_{\alpha \alpha }^{\pm (0)}(u)+\eta k_{\alpha
\alpha }^{\pm (1)}(u)+\frac{1}{2}\eta ^{2}k_{\alpha \alpha }^{\pm (2)}(u)+%
{\rm o}(\eta ^{3}),\qquad (\alpha =1,2,3).  \label{str.8}
\end{equation}

Substituting these expansions into the matrix operator $V=K^{+}(u)U(u|z)$ we
will get a quasi-classical expansion for its diagonal entries:%
\begin{equation}
V_{\alpha \alpha }(u|z)=1+2\eta V_{\alpha \alpha }^{(1)}(u|z)+4\eta
^{2}V_{\alpha \alpha }^{(2)}(u|z)+{\rm o}(\eta ^{3})  \label{str.9}
\end{equation}%
where the first order terms are very simple%
\begin{eqnarray}
V_{11}^{(1)}(u|z) &=&\frac{1}{2}\left[
k_{11}^{+(1)}(u)k_{11}^{-(0)}(u)+k_{11}^{+(0)}(u)k_{11}^{-(1)}(u)\right]  
\nonumber \\
&&+\sum_{a=1}^{N}\coth (u-z_{a}){\cal H}_{a}+\sum_{b=1}^{N}\coth (u+z_{b})%
{\cal H}_{b},  \nonumber \\
V_{22}^{(1)}(u|z) &=&\frac{1}{2}\left[
k_{22}^{+(1)}(u)k_{22}^{-(0)}(u)+k_{22}^{+(0)}(u)k_{22}^{-(1)}(u)\right] , 
\nonumber \\
V_{33}^{(1)}(u|z) &=&\frac{1}{2}\left[
k_{33}^{+(1)}(u)k_{33}^{-(0)}(u)+k_{33}^{+(0)}(u)k_{33}^{-(1)}(u)\right]  
\nonumber \\
&&-\sum_{a=1}^{N}\coth (u-z_{a}){\cal H}_{a}-\sum_{b=1}^{N}\coth (u+z_{b})%
{\cal H}_{b}  \label{str.10}
\end{eqnarray}%
and the second order terms are very complicated 
\begin{eqnarray}
V_{11}^{(2)}(u|z) &=&\frac{1}{8}\left[
k_{11}^{+(2)}(u)k_{11}^{-(0)}(u)+k_{11}^{+(0)}(u)k_{11}^{-(2)}(u)\right] +%
\frac{1}{4}k_{11}^{+(1)}(u)k_{11}^{-(1)}(u)  \nonumber \\
&&+\sum_{a<b}\left\{ \coth (u-z_{a})\coth (u-z_{b}){\cal H}_{a}\overset{s}{%
\otimes }{\cal H}_{b}+\coth (u+z_{b})\coth (u+z_{a}){\cal H}_{b}\overset{s}{%
\otimes }{\cal H}_{a}\right\}   \nonumber \\
&&+\sum_{a<b}\left\{ {\rm e}^{z_{a}-z_{b}}\frac{{\cal X}_{a}^{-}\overset{s}{%
\otimes }{\cal X}_{b}^{+}-{\cal V}_{a}^{-}\overset{s}{\otimes }{\cal V}%
_{b}^{+}}{\sinh (u-z_{a})\sinh (u-z_{b})}+{\rm e}^{z_{b}-z_{a}}\frac{{\cal X}%
_{b}^{-}\overset{s}{\otimes }{\cal X}_{a}^{+}-{\cal V}_{b}^{-}\overset{s}{%
\otimes }{\cal V}_{a}^{+}}{\sinh (u+z_{b})\sinh (u+z_{a})}\right\}  
\nonumber \\
&&+\sum_{a=1}^{L}\left( \frac{1}{2}{\cal H}_{a}^{2}+\frac{3}{4}\frac{{\cal H}%
_{a}^{2}-{\cal H}_{a}}{\sinh (u-z_{a})^{2}}\right) +\sum_{b=1}^{N}\left( 
\frac{1}{2}{\cal H}_{b}^{2}+\frac{3}{4}\frac{{\cal H}_{b}^{2}-{\cal H}_{b}}{%
\sinh (u+z_{b})^{2}}\right)   \nonumber \\
&&+\frac{1}{2}\left[
k_{11}^{+(0)}(u)k_{11}^{-(1)}(u)+k_{11}^{+(1)}(u)k_{11}^{-(0)}(u)\right]
\left\{ \sum_{a=1}^{N}\coth (u-z_{a}){\cal H}_{a}+\sum_{b=1}^{N}\coth
(u+z_{b}){\cal H}_{b}\right\}   \nonumber \\
&&+\sum_{a=1}^{N}\coth (u-z_{a}){\cal H}_{a}\sum_{b=1}^{N}\coth (u+z_{b})%
{\cal H}_{b}  \nonumber \\
&&-k_{11}^{+(0)}(u)k_{22}^{-(0)}(u)\sum_{a=1}^{N}\frac{{\rm e}^{-u+z_{a}}%
{\cal V}_{a}^{-}}{\sinh (u-z_{a})}\sum_{b=1}^{N}\frac{{\rm e}^{-u-z_{b}}%
{\cal V}_{b}^{+}}{\sinh (u+z_{b})}  \nonumber \\
&&+k_{11}^{+(0)}(u)k_{33}^{-(0)}(u)\sum_{a=1}^{N}\frac{{\rm e}^{-u+z_{a}}%
{\cal X}_{a}^{-}}{\sinh (u-z_{a})}\sum_{b=1}^{N}\frac{{\rm e}^{-u-z_{b}}%
{\cal X}_{b}^{+}}{\sinh (u+z_{b})},  \label{str.11}
\end{eqnarray}%
\begin{eqnarray}
V_{22}^{(2)}(u|z) &=&\frac{1}{8}\left[
k_{22}^{+(2)}(u)k_{22}^{-(0)}(u)+k_{22}^{+(0)}(u)k_{22}^{-(2)}(u)\right] +%
\frac{1}{4}k_{22}^{+(1)}(u)k_{22}^{-(1)}(u)  \nonumber \\
&&-\sum_{a<b}\frac{{\rm e}^{-z_{a}+z_{b}}{\cal V}_{a}^{+}\overset{s}{\otimes 
}{\cal V}_{b}^{-}-{\rm e}^{z_{a}-z_{b}}{\cal V}_{a}^{-}\overset{s}{\otimes }%
{\cal V}_{b}^{+}}{\sinh (u-z_{a})\sinh (u-z_{b})}-\frac{3}{2}\sum_{a=1}^{N}%
\frac{{\cal I}_{a}-{\cal H}_{a}^{2}}{\sinh (u-z_{a})^{2}}  \nonumber \\
&&-\sum_{a<b}\frac{{\rm e}^{-z_{b}+z_{a}}{\cal V}_{b}^{+}\overset{s}{\otimes 
}{\cal V}_{a}^{-}-{\rm e}^{z_{b}-z_{a}}{\cal V}_{b}^{-}\overset{s}{\otimes }%
{\cal V}_{a}^{+}}{\sinh (u+z_{b})\sinh (u+z_{b})}-\frac{3}{2}\sum_{b=1}^{N}%
\frac{{\cal I}_{b}-{\cal H}_{b}^{2}}{\sinh (u+z_{b})^{2}}  \nonumber \\
&&-k_{22}^{+(0)}(u)k_{11}^{-(0)}(u)\sum_{a=1}^{N}\frac{{\rm e}^{u-z_{a}}%
{\cal V}_{a}^{+}}{\sinh (u-z_{a})}\sum_{b=1}^{N}\frac{{\rm e}^{u+z_{b}}{\cal %
V}_{b}^{-}}{\sinh (u+z_{b})}  \nonumber \\
&&+k_{22}^{+(0)}(u)k_{33}^{-(0)}(u)\sum_{a=1}^{N}\frac{{\rm e}^{-u+z_{a}}%
{\cal V}_{a}^{-}}{\sinh (u-z_{a})}\sum_{b=1}^{N}\frac{{\rm e}^{-u-z_{b}}%
{\cal V}_{b}^{+}}{\sinh (u+z_{b})},  \label{str.12}
\end{eqnarray}

\begin{eqnarray}
V_{33}^{(2)}(u|z) &=&\frac{1}{8}\left[
k_{33}^{+(2)}(u)k_{33}^{-(0)}(u)+k_{33}^{+(0)}(u)k_{33}^{-(2)}(u)\right] +%
\frac{1}{4}k_{33}^{+(1)}(u)k_{33}^{-(1)}(u)  \nonumber \\
&&+\sum_{a<b}\left\{ \coth (u-z_{a})\coth (u-z_{b}){\cal H}_{a}\overset{s}{%
\otimes }{\cal H}_{b}+\coth (u+z_{b})\coth (u+z_{a}){\cal H}_{b}\overset{s}{%
\otimes }{\cal H}_{a}\right\}  \nonumber \\
&&+\sum_{a<b}\left\{ {\rm e}^{-z_{a}+z_{b}}\frac{{\cal X}_{a}^{+}\overset{s}{%
\otimes }{\cal X}_{b}^{-}+{\cal V}_{a}^{+}\overset{s}{\otimes }{\cal V}%
_{b}^{-}}{\sinh (u-z_{a})\sinh (u-z_{b})}+{\rm e}^{-z_{b}+z_{a}}\frac{{\cal X%
}_{b}^{+}\overset{s}{\otimes }{\cal X}_{a}^{-}+{\cal V}_{b}^{+}\overset{s}{%
\otimes }{\cal V}_{a}^{-}}{\sinh (u+z_{b})\sinh (u+z_{a})}\right\}  \nonumber
\\
&&+\sum_{a=1}^{N}\left( \frac{1}{2}{\cal H}_{a}^{2}+\frac{3}{4}\frac{{\cal H}%
_{a}^{2}+{\cal H}_{a}}{\sinh (u-z_{a})^{2}}\right) +\sum_{b=1}^{N}\left( 
\frac{1}{2}{\cal H}_{b}^{2}+\frac{3}{4}\frac{{\cal H}_{b}^{2}+{\cal H}_{b}}{%
\sinh (u+z_{b})^{2}}\right)  \nonumber \\
&&-\frac{1}{2}\left[
k_{33}^{+(0)}(u)k_{33}^{-(1)}(u)+k_{33}^{+(1)}(u)k_{33}^{-(0)}(u)\right]
\left\{ \sum_{a=1}^{N}\coth (u-z_{a}){\cal H}_{a}+\sum_{b=1}^{N}\coth
(u+z_{b}){\cal H}_{b}\right\}  \nonumber \\
&&+\sum_{a=1}^{N}\coth (u-z_{a}){\cal H}_{a}\sum_{b=1}^{N}\coth (u+z_{b})%
{\cal H}_{b}  \nonumber \\
&&+k_{33}^{+(0)}(u)k_{22}^{-(0)}(u)\sum_{a=1}^{N}\frac{{\rm e}^{u-z_{a}}%
{\cal V}_{a}^{+}}{\sinh (u-z_{a})}\sum_{b=1}^{N}\frac{{\rm e}^{u+z_{b}}{\cal %
V}_{b}^{-}}{\sinh (u+z_{b})}  \nonumber \\
&&+k_{33}^{+(0)}(u)k_{11}^{-(0)}(u)\sum_{a=1}^{N}\frac{{\rm e}^{u-z_{a}}%
{\cal X}_{a}^{+}}{\sinh (u-z_{a})}\sum_{b=1}^{N}\frac{{\rm e}^{u+z_{b}}{\cal %
X}_{b}^{-}}{\sinh (u+z_{b})}.  \label{str.13}
\end{eqnarray}%
Here we have used the identities $k_{\alpha \alpha }^{+(0)}(u)k_{\alpha
\alpha }^{-(0)}(u)=1,\ \alpha =1,2,3.$, which hold for the three boundary
solution due to the isomorphism (\ref{mod.8}).

Now, the quasi-classical expansion for the double-row transfer matrix (\ref%
{mod.11}) has the form%
\begin{eqnarray}
\tau (u|z) &=&V_{11}(u|z)-V_{22}(u|z)+V_{33}(u|z)  \nonumber \\
&=&1+8\eta ^{2}\tau ^{(2)}(u|z)+{\rm o}(\eta ^{3}).  \label{str.14}
\end{eqnarray}%
where we notice that the first order terms in $\eta $ are canceled.

The residue procedure of $\tau ^{(2)}(u|z)$ at the point $u=z_{a}$ results
in the following generalized Gaudin Hamiltonians

\begin{eqnarray}
G_{a} &=&\frac{1}{4}\left(
k_{11}^{+(0)}(z_{a})k_{11}^{-(1)}(z_{a})+k_{11}^{+(1)}(z_{a})k_{11}^{-(0)}(z_{a})-k_{33}^{+(0)}(z_{a})k_{33}^{-(1)}(z_{a})-k_{33}^{+(1)}(z_{a})k_{33}^{-(0)}(z_{a})\right) 
{\cal H}_{a}  \nonumber \\
&&+\sum_{b=1}^{N}\frac{1}{\sinh (z_{a}-z_{b})}\left\{ \cosh (z_{a}-z_{b})%
{\cal H}_{a}\overset{s}{\otimes }{\cal H}_{b}+\frac{1}{2}\ ({\rm e}%
^{z_{a}-z_{b}}{\cal X}_{a}^{-}\overset{s}{\otimes }{\cal X}_{b}^{+}+{\rm e}%
^{z_{b}-z_{a}}{\cal X}_{a}^{+}\overset{s}{\otimes }{\cal X}_{b}^{-})\right. 
\nonumber \\
&&+\left. \ {\rm e}^{z_{b}-z_{a}}{\cal V}_{a}^{+}\overset{s}{\otimes }{\cal V%
}_{b}^{-}-{\rm e}^{z_{a}-z_{b}}{\cal V}_{a}^{-}\overset{s}{\otimes }{\cal V}%
_{b}^{+}\right\}  \nonumber \\
&&+\frac{1}{2}\sum_{b=1}^{N}\frac{1}{\sinh (z_{a}+z_{b})}\left\{ \left(
k_{11}^{+(0)}(z_{a})k_{11}^{-(0)}(z_{a})+k_{33}^{+(0)}(z_{a})k_{33}^{-(0)}(z_{a})\right) \cosh (z_{a}+z_{b})%
{\cal H}_{a}\overset{s}{\otimes }{\cal H}_{b}\right.  \nonumber \\
&&+k_{11}^{+(0)}(z_{a})k_{33}^{-(0)}(z_{a})\ {\rm e}^{-z_{a}-z_{b}}{\cal X}%
_{a}^{-}\overset{s}{\otimes }{\cal X}%
_{b}^{+}+k_{33}^{+(0)}(z_{a})k_{11}^{-(0)}(z_{a})\ {\rm e}^{z_{a}+z_{b}}%
{\cal X}_{a}^{+}\overset{s}{\otimes }{\cal X}_{b}^{-}  \nonumber \\
&&-\left(
k_{11}^{+(0)}(z_{a})k_{22}^{-(0)}(z_{a})+k_{22}^{+(0)}(z_{a})k_{33}^{-(0)}(z_{a})\right) \ 
{\rm e}^{-z_{a}-z_{b}}{\cal V}_{a}^{-}\overset{s}{\otimes }{\cal V}_{b}^{+} 
\nonumber \\
&&+\left. \left(
k_{22}^{+(0)}(z_{a})k_{11}^{-(0)}(z_{a})+k_{33}^{+(0)}(z_{a})k_{22}^{-(0)}(z_{a})\right) \ 
{\rm e}^{z_{a}+z_{b}}{\cal V}_{a}^{+}\overset{s}{\otimes }{\cal V}%
_{b}^{-}\right\}  \label{str.15}
\end{eqnarray}%
which satisfy the following integrable conditions 
\begin{equation}
\sum_{a=1}^{N}G_{a}=0,\quad \frac{\partial G_{a}}{\partial z_{b}}=\frac{%
\partial G_{b}}{\partial z_{a}},\quad \left[ G_{a},G_{b}\right] =0,\qquad
\forall a,b.  \label{str.16}
\end{equation}%
It means that we have derived $N-1$ independents non-local \ integrable
Hamiltonians for each of the three boundary solutions ($1,M$), ($F^{-},G^{-}$%
) and ($F^{+},G^{+}$).

Let us look at the right hand side term of (\ref{str.15}): The first term
represents the diagonal boundary term, the first sum is nothing else but the
sum of classical $r$-matrices $\sum_{b\neq a}r_{ab}(z_{z}-z_{b})$, as it
could be expected and the second sum is also a sum of classical $r$ matrices
but with the boundary effects. This difference emerges due to the breaking
the symmetry of the double-row structure when we choose the point of a
particular row to take the residue. Therefore, we have the following
generalized form 
\begin{equation}
G_{a}=\text{({\rm b.t.)}}_{a}+\sum_{b\neq a}r_{ab}(z_{z}-z_{b})+\sum_{b\neq
a}r_{ab}^{\prime }(z_{z}+z_{b})  \label{str.17}
\end{equation}%
where ({\rm b.t.)}$_{a}$ is the boundary term at the site $a$ (with residue
at $u=z_{a})$, $r_{ab}(z_{z}-z_{b})$ is the "bare" classical $r$-matrix and $%
r_{ab}^{\prime }(z_{z}+z_{b})$ is the classical $r$-matrices "dressed " with
the boundaries.

In the next section we will use the data of the boundary algebraic Bethe
ansatz in order to find the exact spectrum and eigenvectors for these
Hamiltonians.

Next, let us write the entries of the monodromy matrix ${\cal U}(u|z)$, up
to their classical contribution:%
\begin{eqnarray}
{\cal B}_{1}(u|z) &=&-2\eta \sum_{a=1}^{N}\left( \frac{k_{22}^{-(0)}(u){\rm e%
}^{-u+z_{a}}}{\sinh (u-z_{a})}+\frac{k_{11}^{-(0)}(u){\rm e}^{u+z_{a}}}{%
\sinh (u+z_{a})}\right) {\cal V}_{a}^{-}+{\rm o}(\eta ^{2})\doteq -2\eta 
{\cal V}^{-}(u|z)+{\rm o}(\eta ^{2})  \nonumber \\
{\cal B}_{2}(u|z) &=&2\eta \sum_{a=1}^{N}\left( \frac{k_{33}^{-(0)}(u){\rm e}%
^{-u+z_{a}}}{\sinh (u-z_{a})}+\frac{k_{11}^{-(0)}(u){\rm e}^{u+z_{a}}}{\sinh
(u+z_{a})}\right) {\cal X}_{a}^{-}+{\rm o}(\eta ^{2})\doteq 2\eta {\cal X}%
^{-}(u|z)+{\rm o}(\eta ^{2})  \nonumber \\
{\cal B}_{3}(u|z) &=&2\eta \sum_{a=1}^{N}\left( \frac{k_{33}^{-(0)}(u){\rm e}%
^{-u+z_{a}}}{\sinh (u-z_{a})}+\frac{k_{22}^{-(0)}(u){\rm e}^{u+z_{a}}}{\sinh
(u+z_{a})}\right) {\cal V}_{a}^{-}+{\rm o}(\eta ^{2})\doteq 2\eta
k_{11}^{+(0)}(u){\cal V}^{-}(u|z)+{\rm o}(\eta ^{2})  \nonumber \\
&&  \label{str.18}
\end{eqnarray}%
for the creation operators and%
\begin{eqnarray}
{\cal C}_{1}(u|z) &=&2\eta \sum_{a=1}^{N}\left( \frac{k_{11}^{-(0)}(u){\rm e}%
^{u-z_{a}}}{\sinh (u-z_{a})}+\frac{k_{22}^{-(0)}(u){\rm e}^{-u-z_{a}}}{\sinh
(u+z_{a})}\right) {\cal V}_{a}^{+}+{\rm o}(\eta ^{2})\doteq 2\eta {\cal V}%
^{+}(u|z)+{\rm o}(\eta ^{2})  \nonumber \\
{\cal C}_{2}(u|z) &=&2\eta \sum_{a=1}^{N}\left( \frac{k_{11}^{-(0)}(u){\rm e}%
^{u-z_{a}}}{\sinh (u-z_{a})}+\frac{k_{33}^{-(0)}(u){\rm e}^{-u-z_{a}}}{\sinh
(u+z_{a})}\right) {\cal X}_{a}^{+}+{\rm o}(\eta ^{2})\doteq 2\eta {\cal X}%
^{+}(u|z)+{\rm o}(\eta ^{2})  \nonumber \\
{\cal C}_{3}(u|z) &=&2\eta \sum_{a=1}^{N}\left( \frac{k_{22}^{-(0)}(u){\rm e}%
^{u-z_{a}}}{\sinh (u-z_{a})}+\frac{k_{33}^{-(0)}(u){\rm e}^{-u-z_{a}}}{\sinh
(u+z_{a})}\right) {\cal V}_{a}^{+}+{\rm o}(\eta ^{2})\doteq 2\eta
k_{11}^{+(0)}(u){\cal V}^{+}(u|z)+{\rm o}(\eta ^{2})  \nonumber \\
&&  \label{str.19}
\end{eqnarray}%
for the annihilation operators. \ For the diagonal entries of ${\cal U}(u|z)$
we have: 
\begin{eqnarray}
{\cal D}_{1}(u|z) &=&k_{11}^{-(0)}(u)+2\eta {\cal H}(u|z)+{\rm o}(\eta
^{2}),\quad  \nonumber \\
{\cal D}_{2}(u|z) &=&k_{22}^{-(0)}(u)+\eta \ k_{22}^{-(1)}(u)+{\rm o}(\eta
^{2}),\quad  \nonumber \\
{\cal D}_{3}(u|z) &=&k_{33}^{-(0)}(u)+2\eta {\cal H}(-u|z)+{\rm o}(\eta ^{2})
\label{str.20}
\end{eqnarray}%
where%
\begin{equation}
{\cal H}(u|z)=k_{11}^{-(0)}(u)\sum_{a=1}^{N}\left( \cosh (u-z_{a}\right)
+\cosh (u+z_{a}){\cal H}_{a}+\frac{1}{2}k_{11}^{-(1)}(u).  \label{str.21}
\end{equation}%
Here we have used the relations $k_{33}^{-(0)}(u)=k_{11}^{-(0)}(-u),\
k_{22}^{-(0)}(u)=k_{22}^{-(0)}(-u)$ and \ $%
k_{33}^{-(1)}(u)=-k_{11}^{-(1)}(-u)$.

Note that with these expression we are defining the \ generators for a
"double" $osp(1|2)$ Lie superalgebra. The defining relations of this algebra
can be obtained from the quasi-classical limit of (\ref{baba.11a}) or,
equivalently, from the quasi-classical limit of the commutation relations of
the appendix A.

\section{The off-shell Gaudin equation}

The off-shell Gaudin equation is defined from the residue of the
quasi-classical limit of the off-shell Bethe ansatz equation (\ref{baba.17}%
). In this limit we first consider the quasi-classical expansions of the
states (\ref{baba.12}), (\ref{baba.19}) and (\ref{baba.22}) in terms of the
operators (\ref{str.18}) 
\begin{equation}
\Psi _{n}(u_{1},...,u_{n}|z)=(-2\eta )^{n}\Phi _{n}(u_{1},...,u_{n}|z)+{\rm o%
}(\eta ^{n+1}),  \label{off.1}
\end{equation}%
\begin{eqnarray}
\Psi _{n-1}^{j} &=&2(-2\eta )^{n+1}(-1)^{j}\left\{ \frac{1}{2}%
[k_{11}^{+(0)}(u)+k_{11}^{+(0)}(u)k_{22}^{+(0)}(u)]\frac{{\rm e}^{-u+u_{j}}}{%
\sinh (u-u_{j})}\right.  \nonumber \\
&&+\left. \frac{1}{2}[k_{22}^{+(0)}(u)+k_{11}^{+(0)}(u)k_{33}^{+(0)}(u)]%
\frac{{\rm e}^{u+u_{j}}}{\sinh (u+u_{j})}\right\} {\cal V}^{-}(u|z)\Phi
_{n-1}(\overset{\vee }{u_{j}}|z)+{\rm o}(\eta ^{n+2})  \label{off.2}
\end{eqnarray}%
\begin{equation}
\Psi _{n-2}^{lj}=(-2\eta )^{n-1}(-1)^{l+j}{\cal X}^{-}(u|z)\Phi _{n-2}(%
\overset{\vee }{u_{l}},\overset{\vee }{u_{j}}|z)+{\rm o}(\eta ^{n})
\label{off.3}
\end{equation}%
where the quasi-classical Bethe state $\Phi _{n}(u_{1},...,u_{n}|z)$ is
given by the recurrence formula 
\begin{eqnarray}
\Phi _{n}(u_{1},...,u_{n}|z) &=&{\cal V}^{-}(u_{1}|z)\Phi
_{n-1}(u_{2},...,u_{n}|z)  \nonumber \\
&&-{\cal X}^{-}(u_{1}|z)\sum_{j=2}^{n}\frac{(-)^{j}{\rm e}^{-u_{1}+u_{j}}}{%
\sinh (u_{1}-u_{j})}k_{11}^{-(0)}(u_{j})\Phi _{n-2}(\overset{\wedge }{u}%
_{j}|z)  \nonumber \\
&&+{\cal X}^{-}(u_{1}|z)\sum_{j=2}^{n}\frac{(-)^{j}{\rm e}^{-u_{1}-u_{j}}}{%
\sinh (u_{1}+u_{j})}k_{22}^{-(0)}(u_{j})\Phi _{n-2}(\overset{\wedge }{u}%
_{j}|z)  \label{off.4}
\end{eqnarray}%
with $\Phi _{0}=\left\vert 0\right\rangle $ and $\Phi _{1}(u_{1}|z)={\cal V}%
^{-}(u_{1}|z)\Phi _{0}$.

Next we consider the quasi-classical expansions of the c-numbers functions (%
\ref{baba.18}), (\ref{baba.20}) and (\ref{baba.23})%
\begin{eqnarray}
\Lambda _{n} &=&1+2(-2\eta )^{2}\Lambda _{n}^{(2)}+{\rm o}(\eta ^{3}) 
\nonumber \\
{\cal F}_{j}^{(n-1)} &=&(-2\eta )k_{11}^{-(0)}(u_{j})\ \ f_{j}+{\rm o}(\eta
^{2})  \nonumber \\
{\cal F}_{lj}^{(n-2)} &=&2(-2\eta )^{3}\ \ f_{lj}+{\rm o}(\eta ^{4}),
\label{off.5}
\end{eqnarray}%
where%
\begin{eqnarray}
\Lambda _{n}^{(2)} &=&N+\frac{n}{2}+\frac{1}{8}\left(
k_{11}^{+(1)}(u)k_{11}^{-(1)}(u)-k_{22}^{+(1)}(u)k_{22}^{-(1)}(u)+k_{33}^{+(1)}(u)k_{33}^{-(1)}(u)\right)
\nonumber \\
&&+\frac{1}{16}\left\{
k_{11}^{+(0)}(u)k_{11}^{-(2)}(u)-k_{22}^{+(0)}(u)k_{22}^{-(2)}(u)+k_{33}^{+(0)}(u)k_{33}^{-(2)}(u)\right.
\nonumber \\
&&+\left.
k_{11}^{+(2)}(u)k_{11}^{-(0)}(u)-k_{22}^{+(2)}(u)k_{22}^{-(0)}(u)+k_{33}^{+(2)}(u)k_{33}^{-(0)}(u)\right\}
\nonumber \\
&&+\left\{ \frac{1}{2}\left(
4k_{11}^{+(0)}(u)k_{33}^{-(0)}(u)-k_{11}^{+(0)}(u)k_{22}^{-(0)}(u)-k_{33}^{+(0)}(u)k_{22}^{-(0)}(u)\right) 
\frac{{\rm e}^{2u}}{\sinh (2u)}\right.  \nonumber \\
&&+\left. \frac{1}{4}\left(
k_{11}^{+(0)}(u)k_{11}^{-(1)}(u)+k_{11}^{+(1)}(u)k_{11}^{-(0)}(u)-k_{33}^{+(0)}(u)k_{33}^{-(1)}(u)-k_{33}^{+(1)}(u)k_{33}^{-(0)}(u)\right) \right\}
\nonumber \\
&&\times \left( \sum_{a=1}^{N}[\coth (u-z_{a})+\coth
(u+z_{a})]-\sum_{j=1}^{n}[\coth (u-u_{j})+\coth (u+u_{j})]\right)  \nonumber
\\
&&+\sum_{a=1}^{N}\left( \coth (u-z_{a})\coth (u+z_{a})+\frac{3}{4}\frac{1}{%
\sinh (u-z_{a})^{2}}+\frac{3}{4}\frac{1}{\sinh (u+z_{a})^{2}}\right) 
\nonumber \\
&&+\sum_{j=1}^{n}\left( \coth (u-u_{j})\coth (u+u_{j})+\frac{1}{4}\coth
(u-u_{j})^{2}+\frac{1}{4}\coth (u+u_{j})^{2}\right)  \nonumber \\
&&+\sum_{a\,<b=1}^{N}(\coth (u-z_{a})\coth (u+z_{a}))(\coth (u-z_{b})\coth
(u+z_{b}))  \nonumber \\
&&+\sum_{i\,<j=1}^{n}(\coth (u-u_{i})\coth (u+u_{i}))(\coth (u-u_{j})\coth
(u+u_{j}))  \nonumber \\
&&-\sum_{a=1}^{N}(\coth (u-z_{a})+\coth (u+z_{a}))\sum_{j=1}^{n}(\coth
(u-u_{j})+\coth (u+u_{j}))  \label{off.6}
\end{eqnarray}

\begin{eqnarray}
f_{j} &=&-\sum_{a=1}^{N}(\coth (u_{j}-z_{a})+\coth
(u_{j}+z_{a}))+\sum_{k\neq j}^{n}(\coth (u_{j}-u_{k})+\coth (u_{j}+u_{k})) 
\nonumber \\
&&-\frac{1}{2}\left( \frac{k_{11}^{-(1)}(u_{j})}{k_{11}^{-(0)}(u_{j})}%
+\theta ^{(1)}(u_{j})\frac{k_{22}^{-(0)}(u_{j})}{k_{11}^{-(0)}(u_{j})}%
\right) +\left( \frac{{\rm e}^{2u_{j}}}{\sinh (2u_{j})}-\frac{1}{2}\frac{%
k_{22}^{-(1)}(u_{j})}{k_{11}^{-(0)}(u_{j})}\right) \theta ^{(0)}(u_{j})
\label{off.7}
\end{eqnarray}%
Note that in these relations we have expanded (\ref{baba.21}) as $\theta
(u)=\theta ^{(0)}(u)+\eta \theta ^{(1)}(u)+{\rm o}(\eta ^{2})$ and used the
identity $k_{11}^{-(0)}(u_{j})+k_{22}^{-(0)}(u_{j})\theta ^{(0)}(u_{j})=0$
which holds for the three cases.

The contribution of ${\cal F}_{lj}^{(n-2)}$ involves the expansions of the
complicate functions $H_{\alpha j}$ presented in the appendix B. After some
algebraic manipulations we have

\begin{eqnarray}
f_{lj} &=&-\frac{\sinh (2u)}{\sinh (u_{j}+u_{l})\sinh (u_{l}-u_{j})}\left\{ 
\frac{\sinh (2u_{j})}{\sinh (u-u_{l})\sinh (u+u_{l})}\frac{%
k_{11}^{+(0)}(u_{l})+k_{22}^{+(0)}(u_{l})}{%
k_{11}^{+(0)}(u_{j})+k_{22}^{+(0)}(u_{j})}[k_{11}^{-(0)}(u_{l})f_{l}]\right.
\nonumber \\
&&+\left. \frac{\sinh (2u_{l})}{\sinh (u-u_{j})\sinh (u+u_{j})}\frac{%
k_{11}^{+(0)}(u_{j})+k_{22}^{+(0)}(u_{j})}{%
k_{11}^{+(0)}(u_{l})+k_{22}^{+(0)}(u_{l})}[k_{11}^{-(0)}(u_{j})f_{j}]\right\}
\label{off.8}
\end{eqnarray}

Substituting these expressions into the (\ref{baba.17}) and comparing the
coefficients of the terms $2(-2\eta )^{n+2}$ we get the first non-trivial
consequence for the quasi-classical limit of the \ {\small OSBAE}: 
\begin{eqnarray}
\tau ^{(2)}(u|z)\ \Phi _{n}(u_{1},...,u_{n}|z) &=&\Lambda _{n}^{(2)}\ \Phi
_{n}(u_{1},...,u_{n}|z)+\sum_{j=1}^{n}(-1)^{j}\ \frac{{\rm e}^{u_{j}-u}}{%
\sinh (u-u_{j})}\ [k_{11}^{-(0)}(u_{j})\ f_{j}]\Theta _{1}^{j}  \nonumber \\
&&+\sum_{j=1}^{n}\ (-1)^{j}\frac{{\rm e}^{u_{j}+u}}{\sinh (u+u_{j})}%
[k_{11}^{-(0)}(u_{j})\ f_{j}]\Theta _{2}^{j}.  \label{off.9}
\end{eqnarray}%
Here we have used the identity%
\begin{equation}
\frac{\sinh (2v)}{\sinh (u-v)\sinh (u+v)}=\frac{{\rm e}^{-u+v}}{\sinh (u-v)}-%
\frac{{\rm e}^{-u-v}}{\sinh (u+v)}  \label{off.10}
\end{equation}%
Note that in (\ref{off.9}) the contributions from $\Psi _{(n-1)}^{j}$ and $%
\Psi _{(n-2)}^{lj}$ were combined in order to define two new vector valued
functions%
\begin{eqnarray}
\Theta _{1}^{j} &=&\frac{1}{2}%
[k_{11}^{+(0)}(u)+k_{11}^{+(0)}(u)k_{22}^{+(0)}(u)]{\cal V}^{-}(u|z)\ \Phi
_{n-1}(\ \overset{\wedge }{u}_{j}|z)  \nonumber \\
&&-{\cal X}^{-}(u|z)\sum\begin{Sb} k=1  \\ k\neq j  \end{Sb} 
^{n}(-1)^{k^{^{\prime }}}(\frac{{\rm e}^{u_{j}-u_{k}}}{\sinh (u_{k}-u_{j})}+%
\frac{{\rm e}^{u_{j}+u_{k}}}{\sinh (u_{k}+u_{j})})\frac{%
k_{11}^{+(0)}(u_{j})+k_{22}^{+(0)}(u_{j})}{%
k_{11}^{+(0)}(u_{k})+k_{22}^{+(0)}(u_{k})}\ \Phi _{n-2}(\ \overset{\wedge }{u%
}_{j},\overset{\wedge }{u}_{k}|z)  \nonumber \\
&&  \label{off.11}
\end{eqnarray}%
and 
\begin{eqnarray}
\Theta _{2}^{j} &=&\frac{1}{2}%
[k_{22}^{+(0)}(u)+k_{11}^{+(0)}(u)k_{33}^{+(0)}(u)]{\cal V}^{-}(u|z)\ \Phi
_{n-1}(\ \overset{\wedge }{u}_{j}|z)  \nonumber \\
&&-{\cal X}^{-}(u|z)\sum\begin{Sb} k=1  \\ k\neq j  \end{Sb} 
^{n}(-1)^{k^{^{\prime }}}(\frac{{\rm e}^{u_{j}-u_{k}}}{\sinh (u_{k}-u_{j})}+%
\frac{{\rm e}^{u_{j}+u_{k}}}{\sinh (u_{k}+u_{j})})\frac{%
k_{11}^{+(0)}(u_{j})+k_{22}^{+(0)}(u_{j})}{%
k_{11}^{+(0)}(u_{k})+k_{22}^{+(0)}(u_{k})}\ \Phi _{n-2}(\ \overset{\wedge }{u%
}_{j},\overset{\wedge }{u}_{k}|z)  \nonumber \\
&&  \label{off.12}
\end{eqnarray}%
where $k^{^{\prime }}=k+1\ \ $for$\quad k<j$ \ and $k^{^{\prime }}=k\ $\ for$%
\quad k>j$. \ 

Finally, we have the off-shell Gaudin equation taking the residue of (\ref%
{baba.17}) at the point $u=z_{a}$: 
\begin{eqnarray}
G_{a}\Phi _{n}(u_{1},...,u_{n}|z) &=&g_{a}\Phi
_{n}(u_{1},...,u_{n}|z)+\sum_{j=1}^{n}(-1)^{j}\ \frac{{\rm e}^{u_{j}-z_{a}}}{%
\sinh (z_{a}-u_{j})}[\ k_{11}^{-(0)}(u_{j})\ f_{j}]\phi _{1}^{j}  \nonumber
\\
&&+\sum_{j=1}^{n}\ (-1)^{j}\frac{{\rm e}^{u_{j}+z_{a}}}{\sinh (z_{a}+u_{j})}%
[\ k_{11}^{-(0)}(u_{j})\ f_{j}]\phi _{2}^{j},  \nonumber \\
a &=&1,2,...,N  \label{off.13}
\end{eqnarray}%
where $g_{a}$ is the residue of $\Lambda _{n}^{(2)}$ at the point $u=z_{a}$ 
\begin{eqnarray}
g_{a} &=&\left\{ \frac{1}{2}\left(
4k_{11}^{+(0)}(z_{a})k_{33}^{-(0)}(z_{a})-k_{11}^{+(0)}(z_{a})k_{22}^{-(0)}(z_{a})-k_{33}^{+(0)}(z_{a})k_{22}^{-(0)}(z_{a})\right) 
\frac{{\rm e}^{2z_{a}}}{\sinh (2z_{a})}\right.  \nonumber \\
&&+\frac{1}{4}\left. \left(
k_{11}^{+(0)}(z_{a})k_{11}^{-(1)}(z_{a})+k_{11}^{+(1)}(z_{a})k_{11}^{-(0)}(z_{a})-k_{33}^{+(0)}(z_{a})k_{33}^{-(1)}(z_{a})-k_{33}^{+(1)}(z_{a})k_{33}^{-(0)}(z_{a})\right) \right\}
\nonumber \\
&&+\coth (2z_{a})+\sum_{b\neq a}^{N}(\coth (z_{a}-z_{b})\coth
(z_{a}+z_{b}))-\sum_{j=1}^{n}(\coth (z_{a}-u_{j})+\coth (z_{a}+u_{j})),
\label{off.14}
\end{eqnarray}%
The new functions $\phi _{1(2)}^{l}$ are given by $\phi _{1(2)}^{j}={\rm res}%
_{u=z_{a}}\Theta _{1(2)}^{j}$.

In this way we are arriving to the main result of this paper. The equation (%
\ref{off.13}) governs the Gaudin theory: From its reduction to an eigenvalue
problem we can find the eigenvalues and corresponding eigenvectors for $N-1$
commuting Hamiltonians $G_{a}$ (\ref{str.15}). It means that $g_{a}$ is the
eigenvalue of $G_{a}$ with eigenfunction $\Phi _{n}(u_{1},...,u_{n})$
provided $u_{l}$ are solutions of the equations $f_{j}=0$ . Moreover, as we
will see in the next section, the off-shell Gaudin equation defines the
conditions on the monodromy of the correlations functions in order to be
solutions of the differential equations known as {\small KZ} equations.

\section{The Knizhnik-Zamolodchickov equation}

\bigskip The Knizhnik-Zamolodchikov ({\small KZ)} differential equation 
\begin{equation}
\kappa \frac{\partial \Psi (z)}{\partial z_{i}}=G_{i}(z)\Psi (z),
\label{kz.1}
\end{equation}%
appeared first as a \ holonomic system of differential equations on
conformal blocks in a {\small WZW} model of conformal field theory. Here $%
\Psi (z)$ is a function with values in the tensor product $V_{1}\otimes
\cdots \otimes V_{N}$ of representations of a simple Lie algebra, $\kappa
=k+g$ , where $\kappa $ is the central charge of the model, and $g$ is the
dual Coxeter number of the simple Lie algebra.

One of the remarkable properties of the {\small KZ} system is that the
coefficient functions $G_{i}(z)$ commute and that the form $\omega
=\sum_{i}G_{i}(z)dz_{i}$ is closed \cite{RV}: 
\begin{equation}
\frac{\partial G_{j}}{\partial z_{i}}=\frac{\partial G_{i}}{\partial z_{j}}%
,\qquad \left[ G_{i},G_{j}\right] =0.  \label{kz.2}
\end{equation}%
Indeed, it was indicated in \cite{RV} that the equations (\ref{str.2}) are
not just a flatness condition for the form $\omega $ but that the {\small KZ}
connection is actually a commutative family of connections.

In this section we will identify $G_{i}$ with the bounded $osp(1|2)$ Gaudin
Hamiltonians $G_{a}$ presented in the previous section and show that the
corresponding differential equations (\ref{kz.1}) can be solved via the
boundary off-shell Bethe ansatz method.

Let us now define the vector-valued function $\Psi (z_{1},...,z_{N})$
through multiple contour integrals of the vectors (\ref{off.4}) 
\begin{equation}
\Psi (z_{1},...,z_{N})=\oint \cdots \oint {\cal X}(u|z)\Phi
_{n}(u|z)du_{1}...du_{n},  \label{kz.3}
\end{equation}%
where ${\cal X}$ $(u|z)={\cal X}$ $(u_{1},...,u_{n},z_{1},...,z_{N})$ is a
scalar function which in this stage is still undefined.

We assume that $\Psi (z_{1},...,z_{N})$ is a solution of the equations 
\begin{equation}
\kappa \frac{\partial \Psi (z_{1},...,z_{N})}{\partial z_{a}}=G_{a}\Psi
(z_{1},...,z_{N}),\quad a=1,2,...,N  \label{kz.4}
\end{equation}%
where $G_{a}$ are the Gaudin Hamiltonians (\ref{str.15}) and $\kappa $ is a
constant.

Substituting (\ref{kz.3}) into (\ref{kz.4}) we have 
\begin{equation}
\kappa \frac{\partial \Psi (z_{1},...,z_{N})}{\partial z_{a}}=\oint \left\{
\kappa \frac{\partial {\cal X}(u|z)}{\partial z_{a}}\Phi _{n}(u|z)+\kappa 
{\cal X}(u|z)\frac{\partial \Phi _{n}(u|z)}{\partial z_{a}}\right\} du,
\label{kz.5}
\end{equation}%
where we are using a compact notation $\oint \left\{ \circ \right\} du=\oint
\ldots \oint \left\{ \circ \right\} $\ $du_{1}\cdots du_{n}.$

Using the quasi-classical limit of the commutation relations one can derive
the following non-trivial identity 
\begin{equation}
\frac{\partial \Phi _{n}}{\partial z_{a}}=\sum_{l=1}^{n}(-)^{l}\frac{%
\partial }{\partial u_{l}}\left( \frac{{\rm e}^{-u_{l}+z_{a}}\phi _{1}^{l}}{%
\sinh (u_{l}-z_{a})}+\frac{{\rm e}^{u_{l}+z_{a}}\phi _{2}^{l}}{\sinh
(u_{l}+z_{a})}\right)  \label{kz.6}
\end{equation}%
which allows us write (\ref{kz.5}) in the form%
\begin{eqnarray}
\kappa \frac{\partial \Psi }{\partial z_{a}} &=&\oint \left\{ \kappa \frac{%
\partial {\cal X}(u|z)}{\partial z_{a}}\Phi
_{n}(u|z)+\sum_{l=1}^{n}(-)^{l}\kappa \frac{\partial {\cal X}(u|z)}{\partial
u_{l}}[\frac{{\rm e}^{u_{l}-z_{a}}\phi _{1}^{l}}{\sinh (z_{a}-u_{l})}+\frac{%
{\rm e}^{u_{l}+z_{a}}\phi _{2}^{l}}{\sinh (z_{a}+u_{l})}]\right\} du 
\nonumber \\
&&-\kappa \sum_{l=1}^{n}(-)^{l}\oint \frac{\partial }{\partial u_{l}}\left( 
{\cal X}(u|z)[\frac{{\rm e}^{u_{l}-z_{a}}\phi _{1}^{l}}{\sinh (z_{a}-u_{l})}+%
\frac{{\rm e}^{u_{l}+za}\phi _{2}^{l}}{\sinh (z_{a}+u_{l})}]\right) du.
\label{kz.7}
\end{eqnarray}%
It is evident that the last term of (\ref{kz.7}) is vanishes, because the
contours are closed. Moreover, if the scalar function ${\cal X}(u|z)$
satisfies the following differential equations 
\begin{equation}
\kappa \frac{\partial {\cal X}(u|z)}{\partial z_{a}}=g_{a}{\cal X}%
(u|z),\qquad \kappa \frac{\partial {\cal X}(u|z)}{\partial u_{j}}%
=[k_{11}^{-(0)}(u_{j})f_{j}]{\cal X}(u|z),  \label{kz.8}
\end{equation}%
we are recovering the off-shell Gaudin equation (\ref{off.13}) from the
first term in (\ref{kz.7}).

Next we can solve (\ref{kz.8}) in order to find the function ${\cal X}(u|z)$
which determines the monodromy of (\ref{kz.3}) as solution of the
trigonometric {\small KZ} equation.

Summarizing, let us consider the main results for the three boundary
solutions ($1,M$), ($F^{+},G^{+}$) and ($F^{-},G^{-}$)

\subsection{The (1,M) Solution}

The data for this case are%
\begin{eqnarray}
k_{11}^{-(0)}(u) &=&k_{22}^{-(0)}(u)=k_{33}^{-(0)}(u)=1,\
k_{11}^{-(1)}(u)=k_{22}^{-(1)}(u)=k_{33}^{-(1)}(u)=0  \nonumber \\
k_{11}^{+(0)}(u) &=&k_{22}^{+(0)}(u)=k_{33}^{+(0)}(u)=1,\
k_{33}^{+(1)}(u)=-k_{11}^{+(1)}(u)=2,\ k_{22}^{+(1)}(u)=0  \nonumber \\
\theta ^{(0)}(u) &=&-1,\ \theta ^{(1)}(u)=-2\frac{{\rm e}^{2u}}{\sinh (2u)}
\label{K1.1}
\end{eqnarray}%
Substituting (\ref{K1.1}) into (\ref{str.15}), (\ref{off.4}), (\ref{off.14})
and (\ref{off.7}) we have the following generalized Gaudin Hamiltonians

\begin{eqnarray}
G_{a} &=&-{\cal H}_{a}+\sum_{b\neq a}^{N}\frac{1}{\sinh (z_{a}-z_{b})}%
\left\{ \cosh (z_{a}-z_{b}){\cal H}_{a}\overset{s}{\otimes }{\cal H}_{b}+%
\frac{1}{2}\left( {\rm e}^{-z_{a}+z_{b}}{\cal X}_{a}^{+}\overset{s}{\otimes }%
{\cal X}_{b}^{-}\right. \right.  \nonumber \\
&&+\left. {\rm e}^{z_{a}-z_{b}}{\cal X}_{a}^{-}\overset{s}{\otimes }{\cal X}%
_{b}^{+}\right) +\left. {\rm e}^{-z_{a}+z_{b}}{\cal V}_{a}^{+}\overset{s}{%
\otimes }{\cal V}_{b}^{-}-{\rm e}^{z_{a}-z_{b}}{\cal V}_{a}^{-}\overset{s}{%
\otimes }{\cal V}_{b}^{+}\right\}  \nonumber \\
&&+\sum_{b=1}^{N}\frac{1}{\sinh (z_{a}+z_{b})}\left\{ \cosh (z_{a}+z_{b})%
{\cal H}_{a}\overset{s}{\otimes }{\cal H}_{b}+\frac{1}{2}\left( {\rm e}%
^{z_{a}+z_{b}}{\cal X}_{a}^{+}\overset{s}{\otimes }{\cal X}_{b}^{-}\right.
\right.  \nonumber \\
&&+\left. {\rm e}^{-z_{a}-z_{b}}{\cal X}_{a}^{-}\overset{s}{\otimes }{\cal X}%
_{b}^{+}\right) +\left. {\rm e}^{z_{a}+z_{b}}{\cal V}_{a}^{+}\overset{s}{%
\otimes }{\cal V}_{b}^{-}-{\rm e}^{-z_{a}-z_{b}}{\cal V}_{a}^{-}\overset{s}{%
\otimes }{\cal V}_{b}^{+}\right\}  \nonumber \\
a &=&1,2,...,N.  \label{K1.2}
\end{eqnarray}%
with eigenfunctions 
\begin{eqnarray}
\Phi _{n}(u_{1},...,u_{n}|z) &=&{\cal V}^{-}(u_{1}|z)\Phi
_{n-1}(u_{2},...,u_{n}|z)  \nonumber \\
&&-{\cal X}^{-}(u_{1}|z)\sum_{j=2}^{n}\frac{(-)^{j}{\rm e}^{-u_{1}+u_{j}}}{%
\sinh (u_{1}-u_{j})}\Phi _{n-2}(\overset{\wedge }{u}_{j}|z)  \nonumber \\
&&+{\cal X}^{-}(u_{1}|z)\sum_{j=2}^{n}\frac{(-)^{j}{\rm e}^{-u_{1}-u_{j}}}{%
\sinh (u_{1}+u_{j})}\Phi _{n-2}(\overset{\wedge }{u}_{j}|z)  \label{K1.3}
\end{eqnarray}%
and eigenvalues

\begin{equation}
g_{a}=2\coth (2z_{a})+\sum_{b\neq a=1}^{N}[\coth (z_{a}-z_{b})+\coth
(z_{a}+z_{b})]-\sum_{j=1}^{n}[\coth (z_{a}-u_{j})+\coth (z_{a}+u_{j})]
\label{K1.4}
\end{equation}%
provided that%
\begin{equation}
\sum_{a=1}^{N}[\coth (u_{j}-z_{a})+\coth (u_{j}+z_{a})]=\sum_{k\neq
j=1}^{n}[\coth (u_{j}-u_{k})+\coth (u_{j}+u_{k})]  \label{K1.5}
\end{equation}%
The corresponding {\small KZ} solution is 
\begin{equation}
\Psi (z_{1},...,z_{N})=\oint \cdots \oint {\cal X}(u|z)\Phi
_{n}(u|z)du_{1}...du_{n},  \label{K1.6}
\end{equation}%
where%
\begin{eqnarray}
{\cal X}(u|z) &=&[\sinh (2z_{a})]^{1/\kappa
}\prod\limits_{a=1}^{N}\prod\limits_{b=a+1}^{N}[\sinh (z_{a}-z_{b})\sinh
(z_{a}+z_{b})]^{1/\kappa }  \nonumber \\
&&\times \prod\limits_{j=1}^{n}\prod\limits_{k=j+1}^{n}[\sinh
(u_{j}-u_{k})\sinh (u_{j}+u_{k})]^{1/\kappa
}\prod\limits_{a=1}^{N}\prod\limits_{j=1}^{n}[\sinh (z_{a}-u_{j})\sinh
(z_{a}+u_{j})]^{-1/\kappa }  \nonumber \\
&&  \label{K1.7}
\end{eqnarray}%
is the solution of the equations (\ref{kz.8}).

Here we note that the $K$-matrix solution ($1,M$) is a quantum-algebra
invariant solution. Comparing (\ref{K1.2}) with (\ref{str.17}) we can see
that $r^{^{\prime }}=r$. It means that there is no boundary contributions on
the bulk \cite{AMN}.

\subsection{The (F$^{\pm }$,G$^{\pm }$) Solutions}

The data for the ($F^{+},G^{+}$) solution are%
\begin{eqnarray}
k_{11}^{-(0)}(u) &=&-{\rm e}^{-2u},\ k_{22}^{-(0)}(u)=1,\ k_{33}^{-(0)}(u)=-%
{\rm e}^{2u}\   \nonumber \\
k_{11}^{-(1)}(u) &=&-3{\rm e}^{-2u}\coth u,\ k_{22}^{-(1)}(u)=0,\
k_{33}^{-(1)}(u)=-3{\rm e}^{2u}\coth u  \nonumber \\
k_{11}^{+(0)}(u) &=&-{\rm e}^{2u},\ k_{22}^{+(0)}(u)=1,\ k_{33}^{+(0)}(u)=-%
{\rm e}^{-2u}  \nonumber \\
k_{11}^{+(1)}(u) &=&-4{\rm e}^{2u}+3{\rm e}^{2u}\coth u,\
k_{22}^{+(1)}(u)=0,\ k_{33}^{+(1)}(u)=4{\rm e}^{-2u}+3{\rm e}^{-2u}\coth u 
\nonumber \\
\theta ^{(0)}(u) &=&{\rm e}^{-2u},\ \theta ^{(1)}(u)={\rm e}^{-2u}\tanh u
\label{K23.1}
\end{eqnarray}%
while for the ($F^{-},G^{-}$) solution we have.%
\begin{eqnarray}
k_{11}^{-(0)}(u) &=&{\rm e}^{-2u},\ k_{22}^{-(0)}(u)=1,\ k_{33}^{-(0)}(u)=%
{\rm e}^{2u}\   \nonumber \\
k_{11}^{-(1)}(u) &=&3{\rm e}^{-2u}\coth u,\ k_{22}^{-(1)}(u)=0,\
k_{33}^{-(1)}(u)=3{\rm e}^{2u}\coth u  \nonumber \\
k_{11}^{+(0)}(u) &=&{\rm e}^{2u},\ k_{22}^{+(0)}(u)=1,\ k_{33}^{+(0)}(u)=%
{\rm e}^{-2u}  \nonumber \\
k_{11}^{+(1)}(u) &=&4{\rm e}^{2u}-3{\rm e}^{2u}\coth u,\
k_{22}^{+(1)}(u)=0,\ k_{33}^{+(1)}(u)=-4{\rm e}^{-2u}-3{\rm e}^{-2u}\coth u 
\nonumber \\
\theta ^{(0)}(u) &=&{\rm e}^{-2u},\ \theta ^{(1)}(u)={\rm e}^{-2u}\tanh u
\label{K23.2}
\end{eqnarray}%
We will present the summary of these two cases in a compact notation.

Substituting (\ref{K23.1}) ((\ref{K23.2})) into (\ref{str.15}), (\ref{off.4}%
), (\ref{off.14}) and (\ref{off.7}) we have

\begin{eqnarray}
G_{a} &=&2{\cal H}_{a}+\sum_{b=1}^{N}\frac{1}{\sinh (z_{a}+z_{b})}\left\{
\cosh (z_{a}+z_{b}){\cal H}_{a}\overset{s}{\otimes }{\cal H}_{b}+\frac{1}{2}%
\left( {\rm e}^{-4z_{a}}{\rm e}^{z_{a}+z_{b}}{\cal X}_{a}^{+}\overset{s}{%
\otimes }{\cal X}_{b}^{-}\right. \right.  \nonumber \\
&&+\left. {\rm e}^{4z_{a}}{\rm e}^{-z_{a}-z_{b}}{\cal X}_{a}^{-}\overset{s}{%
\otimes }{\cal X}_{b}^{+}\right) +\left. \epsilon \left( {\rm e}^{-2z_{a}}%
{\rm e}^{z_{a}+z_{b}}{\cal V}_{a}^{+}\overset{s}{\otimes }{\cal V}_{b}^{-}-%
{\rm e}^{2z_{a}}{\rm e}^{-z_{a}-z_{b}}{\cal V}_{a}^{-}\overset{s}{\otimes }%
{\cal V}_{b}^{+}\right) \right\}  \nonumber \\
&&+\sum_{b\neq a}^{N}\frac{1}{\sinh (z_{a}-z_{b})}\left\{ \cosh (z_{a}-z_{b})%
{\cal H}_{a}\overset{s}{\otimes }{\cal H}_{b}+\frac{1}{2}\left( {\rm e}%
^{-z_{a}+z_{b}}{\cal X}_{a}^{+}\overset{s}{\otimes }{\cal X}_{b}^{-}\right.
\right.  \nonumber \\
&&+\left. {\rm e}^{z_{a}-z_{b}}{\cal X}_{a}^{-}\overset{s}{\otimes }{\cal X}%
_{b}^{+}\right) +\left. {\rm e}^{-z_{a}+z_{b}}{\cal V}_{a}^{+}\overset{s}{%
\otimes }{\cal V}_{b}^{-}-{\rm e}^{z_{a}-z_{b}}{\cal V}_{a}^{-}\overset{s}{%
\otimes }{\cal V}_{b}^{+}\right\}  \label{K23.3}
\end{eqnarray}%
with eigenfunctions 
\begin{eqnarray}
\Phi _{n}(u_{1},...,u_{n}|z) &=&{\cal V}^{-}(u_{1}|z)\Phi
_{n-1}(u_{2},...,u_{n}|z)  \nonumber \\
&&+{\cal X}^{-}(u_{1}|z)\sum_{j=2}^{n}\frac{(-)^{j}{\rm e}^{-u_{1}+u_{j}}}{%
\sinh (u_{1}-u_{j})}[\epsilon {\rm e}^{-2u_{j}}]\Phi _{n-2}(\overset{\wedge }%
{u}_{j}|z)  \nonumber \\
&&+{\cal X}^{-}(u_{1}|z)\sum_{j=2}^{n}\frac{(-)^{j}{\rm e}^{-u_{1}-u_{j}}}{%
\sinh (u_{1}+u_{j})}\Phi _{n-2}(\overset{\wedge }{u}_{j}|z)  \label{K23.4}
\end{eqnarray}%
and eigenvalues%
\begin{eqnarray}
g_{a} &=&\sum_{b\neq a=1}^{N}[\coth (z_{a}-z_{b})+\coth
(z_{a}+z_{b})]-\sum_{j=1}^{n}[\coth (z_{a}-u_{j})+\coth (z_{a}+u_{j})] 
\nonumber \\
&&+2\coth (2z_{a})+\frac{1}{2}(1+\epsilon )\coth (z_{a})+\frac{1}{2}%
(1-\epsilon )\tanh (z_{a})  \label{K23.5}
\end{eqnarray}%
provided that

\begin{eqnarray}
\sum_{a=1}^{N}[\coth (u_{j}-z_{a})+\coth (u_{j}+z_{a})] &=&\sum_{k\neq
j=1}^{n}[\coth (u_{j}-u_{k})+\coth (u_{j}+u_{k})]  \nonumber \\
&&-\frac{1}{2}(1+\epsilon )\coth (u_{j})-\frac{1}{2}(1-\epsilon )\tanh
(u_{j})  \label{K23.6}
\end{eqnarray}%
The KZ solution is 
\begin{equation}
\Psi _{\epsilon }(z_{1},...,z_{N})=\oint \cdots \oint {\cal X}(u|z)\Phi
_{n}(u|z)du_{1}...du_{n},  \label{K23.7}
\end{equation}%
with%
\begin{eqnarray}
{\cal X}(u|z) &=&[\cosh (u_{j})]^{-(1-\epsilon )/2\kappa }[\cosh
(z_{a})]^{(1-\epsilon )/2\kappa }[\sinh (u_{j})]^{-(1+\epsilon )/2\kappa
}[\sinh (z_{a})]^{(1+\epsilon )/2\kappa }[\sinh (2z_{a})]^{1/\kappa } 
\nonumber \\
&&\times \prod\limits_{a=1}^{N}\prod\limits_{b=a+1}^{N}[\sinh
(z_{a}-z_{b})\sinh (z_{a}+z_{b})]^{1/\kappa
}\prod\limits_{j=1}^{n}\prod\limits_{k=j+1}^{n}[\sinh (u_{j}-u_{k})\sinh
(u_{j}+u_{k})]^{1/\kappa }  \nonumber \\
&&\times \prod\limits_{a=1}^{N}\prod\limits_{j=1}^{n}[\sinh
(z_{a}-u_{j})\sinh (z_{a}+u_{j})]^{-1/\kappa }  \label{K23.8}
\end{eqnarray}%
where $\epsilon =1$ for ($F^{-},G^{-}$) case and $\epsilon =-1$ for ($%
F^{+},G^{+}$) case.

\section{Conclusion}

In this paper a detailed analysis of the boundary quantum inverse scattering
method is applied in order to derive the off-shell Bethe ansatz equation (%
\ref{baba.17}) for a graded $19$-vertex model based on the orthosymplectic
Lie superalgebra $osp(1|2)$. The boundaries ares given by three of solutions
of the reflection equation and its dual, related in pairs by the isomorphism
(\ref{mod.8}). The quasi-classical limit of this equation results in the
off-shell Gaudin equation (\ref{off.13}) and we have emphasized its
importance in the resolution of the Gaudin's theory. From this equation are
following the energy spectrum of the Gaudin magnets and the monodromy
function for the integral representation of the correlation function as
solution of the KZ equation.

There are several issues for which this paper could be useful: By the method
presented in this paper one can include the other three 19-vertex models
namely, the twisted $sl(2|1)^{(2)}$ model , the Zamolodchikov-Fateev model \
and the Izergin-Korepin model for which the algebraic Bethe ansatz with
diagonal boundary are known \cite{China, KLS3, KLS4}. We \ also would like
to know how to connect the rational limit of our three trigonometric results
presented in this work can be related with the $osp(1|2)$ conformal field
theory with conformal boundary conditions \cite{Pearce} and finally, should
be really interesting to try extended these results in order to include all
non-diagonal $K$-matrices classified in \cite{LS2}.

\vspace{.35cm}%

{\bf Acknowledgment:} This work was supported in part by Funda\c{c}\~{a}o de
Amparo \`{a} Pesquisa do Estado de S\~{a}o Paulo-{\small FAPESP}-Brasil and
by Conselho Nacional de Desenvolvimento-{\small CNPq}-Brasil.

\newpage%

\appendix{}

\section{The Commutation Relations}

The equation (\ref{baba.11a}) gives us the commutation relations for the
matrix elements of the double-row monodromy matrix which play a fundamental
role in the algebraic Bethe Ansatz. Here we present the commutations
relations and their coefficientes using a compact notation. Recall that the
entries of the operator 
\begin{equation}
{\cal U}(u)=\left( 
\begin{array}{ccc}
{\cal D}_{1}(u) & {\cal B}_{1}(u) & {\cal B}_{2}(u) \\ 
{\cal C}_{1}(u) & {\cal D}_{2}(u) & {\cal B}_{3}(u) \\ 
{\cal C}_{2}(u) & {\cal C}_{3}(u) & {\cal D}_{3}(u)%
\end{array}%
\right) ,  \label{A.1}
\end{equation}%
are given by (\ref{baba.5}).

Substituting (\ref{mod.15}) and (\ref{A.1}) into (\ref{baba.11a}) we get $81$
equations involving products of two matrix elements of ${\cal U}(u)$. These
equations can be manipulated in ordem to put the product of pairs of
operators in the normal ordered form. To do this we shall proceed in the
following way. First we denote by $E[i,j]=0$ the $(i,j)$ component of the
matrix equation (\ref{baba.11a}) and collect them in blocks $%
B[i,j],i=1,...,5,j=i,...,10-i,$ defined by%
\begin{equation}
B[i,j]=\left\{ F_{ij}=E[i,j],\ f_{ij}=E[j,i],\ FF_{ij}=E[10-i,10-j],\
ff_{ij}=E[10-j,10-i]\right\}  \label{A.3}
\end{equation}%
From these blocks we can see that the pair $(F_{ij},f_{ij})$ as well as $%
(FF_{ij},ff_{ij})$ \ can be solved simultaneously.

We introduce the notation \ 
\begin{equation}
D_{i}={\cal D}_{i}(u),\quad d_{i}={\cal D}_{i}(v),\quad B_{i}={\cal B}%
_{i}(u),\quad b_{i}={\cal B}_{i}(v),\quad C_{i}={\cal C}_{i}(u),\quad c_{i}=%
{\cal C}_{i}(v)  \label{A.4}
\end{equation}%
for the operators of the double-row monodromy matrix and%
\begin{equation}
X_{i}=x_{i}(u+v),\quad Y_{i}=y_{i}(u+v),\quad x_{i}=x_{i}(u-v),\quad
y_{i}=y_{i}(u-v),  \label{A.5}
\end{equation}%
for the Boltzmann weigths and%
\begin{equation}
\{Z\}_{ij}=\{z\}_{ij}(v,u),\qquad \{z\}_{ij}=\{z\}_{ij}(u,v),\quad
F_{i}=f_{i}(u),\quad f_{i}=f_{i}(v).  \label{A.6}
\end{equation}%
for the coefficientes of the commutations relations, where 
\begin{equation}
\{Z\}=A,B,C,D,E,X\qquad {\rm and}\qquad \{z\}=a,b,c,d,e,x.
\end{equation}

Taking into account these simplifications, we will indicate the pair $%
(F_{ij},f_{ij})$ or $(FF_{ij},ff_{ij})$ for which the corresponding normal
ordered relations were obtained:

\begin{itemize}
\item $(F_{14},f_{14})$%
\begin{eqnarray}
D_{1}b_{1}
&=&a_{11}b_{1}D_{1}+a_{12}B_{1}d_{1}+a_{13}B_{1}d_{2}+a_{14}B_{2}c_{1}+a_{15}B_{2}c_{3}+a_{16}b_{2}C_{1}
\nonumber \\
C_{1}d_{1}
&=&A_{11}d_{1}C_{1}+A_{12}D_{1}c_{1}+A_{13}D_{2}c_{1}+A_{14}B_{1}c_{2}+A_{15}B_{3}c_{2}+A_{16}b_{1}C_{2}
\label{A.7}
\end{eqnarray}%
where the coefficients are%
\begin{eqnarray}
A_{11} &=&\frac{x_{1}}{x_{2}}\frac{X_{2}}{X_{1}},\quad A_{12}=-F_{1}\frac{%
X_{5}}{X_{1}}-\frac{y_{5}}{x_{2}}\frac{X_{2}}{X_{1}},\quad A_{13}=-\frac{%
X_{5}}{X_{1}},\quad  \nonumber \\
A_{14} &=&\frac{y_{5}}{x_{2}}\frac{X_{6}}{X_{1}},\quad A_{15}=-\frac{X_{7}}{%
X_{1}},\quad A_{16}=-\frac{x_{1}}{x_{2}}\frac{X_{6}}{X_{1}}.  \label{A.8}
\end{eqnarray}

\item $(F_{17},f_{17})$ 
\begin{eqnarray}
D_{1}b_{2}
&=&b_{11}b_{2}D_{1}+b_{12}B_{2}d_{1}+b_{13}B_{2}d_{2}+b_{14}B_{2}d_{3}+b_{15}B_{1}b_{1}+b_{16}B_{1}b_{3}
\nonumber \\
C_{2}d_{1}
&=&B_{11}d_{1}C_{2}+B_{12}D_{1}c_{2}+B_{13}D_{2}c_{2}+B_{14}D_{3}c_{2}+B_{15}C_{1}c_{1}+B_{16}C_{3}c_{1}
\label{A.9}
\end{eqnarray}%
where%
\begin{eqnarray}
B_{11} &=&\frac{x_{1}}{x_{3}}\frac{X_{3}}{X_{1}},\quad B_{12}=F_{1}\frac{%
y_{6}}{x_{3}}\frac{X_{6}}{X_{1}}-F_{2}\frac{X_{7}}{X_{1}}-\frac{y_{7}}{x_{3}}%
\frac{X_{3}}{X_{1}},\quad B_{13}=\frac{y_{6}}{x_{3}}\frac{X_{6}}{X_{1}}-F_{3}%
\frac{X_{7}}{X_{1}},\quad  \nonumber \\
B_{14} &=&-\frac{X_{7}}{X_{1}},\quad B_{15}=-\frac{y_{6}}{x_{3}}\frac{X_{2}}{%
X_{1}},\quad B_{16}=-\frac{X_{5}}{X_{1}}.  \label{A.10}
\end{eqnarray}

\item $(FF_{36},ff_{36})$ 
\begin{eqnarray}
D_{1}b_{3}
&=&x_{11}b_{3}D_{1}+x_{12}b_{1}D_{1}+x_{13}B_{1}d_{1}+x_{14}B_{1}d_{2}+x_{15}B_{1}d_{3}+x_{16}B_{2}c_{1}
\nonumber \\
&&+x_{17}B_{2}c_{3}+x_{18}b_{2}C_{1}  \nonumber \\
C_{3}d_{1}
&=&X_{11}d_{1}C_{3}+X_{12}d_{1}C_{1}+X_{13}D_{1}c_{1}+X_{14}D_{2}c_{1}+X_{15}D_{3}c_{1}+X_{16}B_{1}c_{2}
\nonumber \\
&&+X_{17}B_{3}c_{2}+X_{18}b_{1}C_{2}  \label{A.11}
\end{eqnarray}%
with the following coefficients%
\begin{eqnarray}
X_{11} &=&\frac{x_{2}}{x_{3}}\frac{X_{3}}{X_{2}},\quad X_{12}=\frac{y_{5}}{%
x_{3}}\frac{Y_{6}}{X_{2}}-\frac{y_{6}}{x_{3}}\frac{Y_{5}}{X_{2}}A_{11},\quad
\nonumber \\
X_{13} &=&F_{1}\frac{y_{6}}{x_{3}}\frac{X_{4}}{X_{2}}-F_{2}\frac{X_{6}}{X_{2}%
}-\frac{y_{7}}{x_{3}}\frac{Y_{6}}{X_{2}}-\frac{y_{6}}{x_{3}}\frac{Y_{5}}{%
X_{2}}A_{12}  \nonumber \\
X_{14} &=&-F_{3}\frac{X_{6}}{X_{2}}+\frac{y_{6}}{x_{3}}\frac{X_{4}}{X_{2}}-%
\frac{y_{6}}{x_{3}}\frac{Y_{5}}{X_{2}}A_{13},\quad  \nonumber \\
X_{15} &=&-\frac{X_{6}}{X_{2}},\quad X_{16}=-\frac{y_{7}}{x_{3}}-\frac{y_{6}%
}{x_{3}}\frac{Y_{5}}{X_{2}}A_{14}  \nonumber \\
X_{17} &=&-\frac{y_{6}}{x_{3}}\frac{X_{5}}{X_{2}}-\frac{y_{6}}{x_{3}}\frac{%
Y_{5}}{X_{2}}A_{15},\quad X_{18}=\frac{y_{5}}{x_{3}}-\frac{y_{6}}{x_{3}}%
\frac{Y_{5}}{X_{2}}A_{16}.  \label{A.12}
\end{eqnarray}%
Note that for each pair of equations the corresponding commutation relations
are related by interchanging 
\begin{equation}
u\leftrightarrow v,\qquad D_{i}\leftrightarrow d_{i},\qquad
B_{i}\leftrightarrow c_{i},\qquad C_{i}\leftrightarrow b_{i}  \label{A.13}
\end{equation}

\item $(F_{24},f_{24})$ 
\begin{eqnarray}
C_{1}b_{1}
&=&c_{11}b_{1}C_{1}+c_{12}b_{1}C_{3}+c_{13}B_{1}c_{3}+c_{14}B_{3}c_{3}+c_{15}b_{2}C_{2}+c_{16}d_{1}D_{1}+c_{17}d_{1}D_{2}
\nonumber \\
&&+c_{18}D_{1}d_{1}+c_{19}D_{1}d_{2}+c_{110}D_{2}d_{1}+c_{111}D_{2}d_{2}
\label{A.14}
\end{eqnarray}%
with the following coefficients%
\begin{eqnarray}
c_{11} &=&-\frac{X_{4}}{X_{1}},\quad c_{12}=\frac{x_{5}}{x_{2}}\frac{X_{6}}{%
X_{1}},\quad c_{13}=-\frac{y_{5}}{x_{2}}\frac{X_{6}}{X_{1}},\quad c_{14}=-%
\frac{X_{7}}{X_{1}},\quad c_{15}=\frac{X_{5}}{X_{1}},  \nonumber \\
c_{16} &=&\frac{Y_{5}}{X_{1}}+F_{1}\frac{x_{5}}{x_{2}}\frac{X_{2}}{X_{1}}%
,\quad c_{17}=\frac{x_{5}}{x_{2}}\frac{X_{2}}{X_{1}},\quad c_{18}=-f_{1}(%
\frac{y_{5}}{x_{2}}\frac{X_{2}}{X_{1}}+F_{1}\frac{X_{5}}{X_{1}}),  \nonumber
\\
c_{19} &=&-(\frac{y_{5}}{x_{2}}\frac{X_{2}}{X_{1}}+F_{1}\frac{X_{5}}{X_{1}}%
),\quad c_{110}=-f_{1}\frac{X_{5}}{X_{1}},\quad c_{111}=-\frac{X_{5}}{X_{1}}.
\label{A.15}
\end{eqnarray}%
The pairs $(F_{16},f_{16}),\ (F_{18},f_{18}),\ (FF_{16},ff_{16})$, $%
(FF_{18},ff_{18})$ and $F_{19}$ form a closed set of equations from which we
have derived the following commutation relations%
\begin{eqnarray}
B_{2}b_{1} &=&e_{11}b_{1}B_{2}+e_{12}b_{2}B_{1}+e_{13}b_{2}B_{3},  \nonumber
\\
C_{1}c_{2} &=&E_{11}c_{2}C_{1}+E_{12}c_{1}C_{2}+E_{13}c_{3}C_{2}  \nonumber
\\
B_{3}b_{2} &=&e_{41}b_{2}B_{3}+e_{42}b_{1}B_{2}+e_{43}b_{3}B_{2}  \nonumber
\\
C_{2}c_{3} &=&E_{41}c_{3}C_{2}+E_{42}c_{2}C_{1}+E_{43}c_{2}C_{3}  \nonumber
\\
B_{1}b_{2}
&=&e_{21}b_{2}B_{1}+e_{22}b_{2}B_{3}+e_{23}b_{1}B_{2}+e_{24}b_{3}B_{2} 
\nonumber \\
C_{2}c_{1}
&=&E_{21}c_{1}C_{2}+E_{22}c_{3}C_{2}+E_{23}c_{2}C_{1}+E_{24}c_{2}C3 
\nonumber \\
B_{2}b_{3}
&=&e_{31}b_{3}B_{2}+e_{32}b_{1}B_{2}+e_{33}b_{2}B_{1}+e_{34}b_{2}B_{3} 
\nonumber \\
C_{3}c_{2}
&=&E_{31}c_{2}C_{3}+E_{32}c_{2}C_{1}+E_{33}c_{1}C_{2}+E_{34}c_{3}C_{2} 
\nonumber \\
B_{2}b_{2} &=&b_{2}B_{2},\qquad C_{2}c_{2}=c_{2}C_{2}  \label{A.16}
\end{eqnarray}%
where%
\begin{equation}
e_{11}=\frac{x_{2}}{x_{1}}\frac{X_{2}}{X_{3}},\quad e_{12}=\frac{y_{5}}{x_{1}%
},\quad e_{13}=-\frac{x_{2}}{x_{1}}\frac{X_{6}}{X_{3}}.
\end{equation}%
\begin{equation}
E_{41}=\frac{x_{2}}{x_{1}}\frac{X_{2}}{X_{3}},\quad E_{42}=\frac{x_{2}}{x_{1}%
}\frac{Y_{6}}{X_{3}},\quad E_{43}=\frac{x_{5}}{x_{1}}
\end{equation}%
\begin{eqnarray}
e_{21} &=&\frac{x_{2}}{x_{1}}\frac{X_{2}X_{3}}{X_{2}^{2}+X_{6}Y_{6}},\quad
e_{22}=\frac{x_{5}(x_{1}^{2}+x_{2}^{2}-x_{5}y_{5})}{%
x_{1}(x_{2}^{2}-x_{5}y_{5})}\frac{X_{2}X_{6}}{X_{2}^{2}+X_{6}Y_{6}} 
\nonumber \\
e_{23} &=&-\frac{x_{1}x_{5}}{x_{2}^{2}-x_{5}y_{5}}\frac{X_{6}Y_{6}}{%
X_{2}^{2}+X_{6}Y_{6}}+\frac{x_{5}}{x_{1}}\frac{X_{2}^{2}}{%
X_{2}^{2}+X_{6}Y_{6}},\quad e_{24}=-\frac{x_{1}x_{2}}{x_{2}^{2}-x_{5}y_{5}}%
\frac{X_{3}X_{6}}{X_{2}^{2}+X_{6}Y_{6}}
\end{eqnarray}%
\begin{eqnarray}
e_{31} &=&\frac{x_{1}x_{2}}{x_{2}^{2}-x_{5}y_{5}}\frac{X_{2}X_{3}}{%
X_{2}^{2}+X_{6}Y_{6}},\quad e_{32}=-\frac{%
x_{5}(x_{1}^{2}+x_{2}^{2}-x_{5}y_{5})}{x_{1}(x_{2}^{2}-x_{5}y_{5})}\frac{%
X_{2}Y_{6}}{X_{2}^{2}+X_{6}Y_{6}}  \nonumber \\
e_{33} &=&-\frac{x_{2}}{x_{1}}\frac{X_{3}Y_{6}}{X_{2}^{2}+X_{6}Y_{6}},\quad
e_{34}=\frac{x_{5}}{x_{1}}\frac{X_{6}Y_{6}}{X_{2}^{2}+X_{6}Y_{6}}-\frac{%
x_{1}x_{5}}{x_{2}^{2}-x_{5}y_{5}}\frac{X_{2}^{2}}{X_{2}^{2}+X_{6}Y_{6}}
\end{eqnarray}

\item $(F_{45},f_{45})$
\end{itemize}

\begin{eqnarray}
D_{2}b_{1}
&=&a_{21}b_{1}D_{2}+a_{22}B_{1}d_{1}+a_{23}B_{1}d_{2}+a_{24}B_{3}d_{1}+a_{25}B_{3}d_{2}+a_{26}B_{2}c_{1}
\nonumber \\
&&+a_{27}B_{2}c_{3}+a_{28}b_{2}C_{1}+a_{29}b_{2}C_{3}  \nonumber \\
C_{1}d_{2}
&=&A_{21}d_{2}C_{1}+A_{22}D_{1}c_{1}+A_{23}D_{2}c_{1}+A_{24}D_{1}c_{3}+A_{25}D_{2}c_{3}+A_{26}B_{1}c_{2}
\nonumber \\
&&+A_{27}B_{3}c_{2}+A_{28}b_{1}C_{2}+A_{29}b_{3}C_{2}
\end{eqnarray}

with%
\begin{eqnarray}
a_{21} &=&\frac{x_{4}}{x_{2}}\frac{X_{4}}{X_{2}}-F_{3}\frac{x_{6}}{x_{2}}%
\frac{X_{6}}{X_{2}}-\frac{x_{4}}{x_{2}}\frac{Y_{5}}{X_{2}}A_{13}-\frac{x_{6}%
}{x_{2}}X_{14}  \nonumber \\
a_{22} &=&f_{1}\frac{y_{5}}{x_{2}}\frac{X_{4}}{X_{2}}-(F_{1}+\frac{y_{5}}{%
x_{2}}\frac{Y_{5}}{X_{2}})a_{12}-\frac{x_{4}}{x_{2}}\frac{Y_{5}}{X_{2}}%
A_{11}-\frac{x_{6}}{x_{2}}X_{12}  \nonumber \\
a_{23} &=&\frac{y_{5}}{x_{2}}\frac{X_{4}}{X_{2}}-(F_{1}+\frac{y_{5}}{x_{2}}%
\frac{Y_{5}}{X_{2}})a_{13},\quad a_{24}=f_{1}\frac{X_{6}}{X_{2}}-\frac{x_{6}%
}{x_{2}}X_{11},\quad a_{25}=\frac{X_{6}}{X_{2}}  \nonumber \\
a_{26} &=&-(F_{1}+\frac{y_{5}}{x_{2}}\frac{Y_{5}}{X_{2}})a_{14}-\frac{x_{4}}{%
x_{2}}\frac{Y_{5}}{X_{2}}A_{16}-\frac{x_{6}}{x_{2}}X_{18}  \nonumber \\
a_{27} &=&-\frac{y_{5}}{x_{2}}\frac{X_{5}}{X_{2}}-(F_{1}+\frac{y_{5}}{x_{2}}%
\frac{Y_{5}}{X_{2}})a_{15},\quad a_{28}=-\frac{y_{6}}{x_{2}}-(F_{1}+\frac{%
y_{5}}{x_{2}}\frac{Y_{5}}{X_{2}})a_{16}-\frac{x_{4}}{x_{2}}\frac{Y_{5}}{X_{2}%
}A_{14}-\frac{x_{6}}{x_{2}}X_{16}  \nonumber \\
a_{29} &=&-\frac{x_{4}}{x_{2}}\frac{X_{5}}{X_{2}}-\frac{x_{4}}{x_{2}}\frac{%
Y_{5}}{X_{2}}A_{15}-\frac{x_{6}}{x_{2}}X_{17}
\end{eqnarray}

\begin{itemize}
\item $(F_{15},f_{15})$ 
\begin{eqnarray}
B_{1}b_{1}
&=&e_{01}b_{1}B_{1}+e_{02}b_{2}D_{2}+e_{03}b_{2}D_{1}+e_{04}B_{2}d_{1}+e_{05}B_{2}d_{2}
\nonumber \\
C_{1}c_{1}
&=&E_{01}c_{1}C_{1}+E_{02}d_{2}C_{2}+E_{03}d_{1}C_{2}+E_{04}D_{1}c_{2}+E_{05}D_{2}c_{2}
\end{eqnarray}%
with%
\begin{eqnarray}
e_{01} &=&-\frac{x_{3}x_{4}-x_{6}y_{6}}{x_{1}x_{3}},\quad e_{02}=\frac{%
x_{3}x_{4}-x_{6}y_{6}}{x_{1}x_{3}}\frac{X_{6}}{X_{2}}  \nonumber \\
e_{03} &=&-\frac{x_{3}y_{6}-x_{6}y_{7}}{x_{1}x_{3}}\frac{X_{3}}{X_{2}}+F_{1}%
\frac{x_{3}x_{4}-x_{6}y_{6}}{x_{1}x_{3}}\frac{X_{6}}{X_{2}}  \nonumber \\
e_{04} &=&-\frac{x_{6}}{x_{3}}\frac{X_{3}}{X_{2}}+f_{1}\frac{X_{6}}{X_{2}}%
,\quad e_{05}=\frac{X_{6}}{X_{2}}
\end{eqnarray}%
These commutaion relation play a special role in the construction of the $n$%
-particle states, as mentioned above.

\item $(FF_{26},ff_{26})$ 
\begin{eqnarray}
B_{1}b_{3}
&=&d_{11}b_{3}B_{1}+d_{12}b_{1}B_{1}+d_{13}b_{2}D_{1}+d_{14}b_{2}D_{2}+b_{15}B_{2}d_{1}+d_{16}B_{2}d_{2}+d_{17}B_{2}d_{3}
\nonumber \\
C_{3}c_{1}
&=&D_{11}c_{1}C_{3}+D_{12}c_{1}C_{1}+D_{13}d_{1}C_{2}+D_{14}d_{2}C_{2}+D_{15}D_{1}c_{2}+D_{16}D_{2}c_{2}+D_{17}D_{3}c_{2}
\nonumber \\
&&
\end{eqnarray}%
where%
\begin{eqnarray}
D_{11} &=&\frac{X_{3}}{Y_{5}B_{16}-X_{4}},\quad D_{12}=-\frac{%
(y_{5}Y_{6}+x_{2}Y_{5}B_{15})E_{01}-y_{5}Y_{6}}{x_{2}(Y_{5}B_{16}-X_{4})} 
\nonumber \\
D_{13} &=&-\frac{%
(y_{5}Y_{6}+x_{2}Y_{5}B_{15})E_{03}-f_{1}y_{5}X_{2}+x_{2}Y_{5}B_{11}}{%
x_{2}(Y_{5}B_{16}-X_{4})},\quad  \nonumber \\
D_{14} &=&-\frac{(y_{5}Y_{6}+x_{2}Y_{5}B_{15})E_{02}-y_{5}X_{2}}{%
x_{2}(Y_{5}B_{16}-X_{4})}  \nonumber \\
D_{15} &=&-\frac{%
(y_{5}Y_{6}+x_{2}Y_{5}B_{15})E_{04}+F_{1}y_{5}X_{2}+F_{2}x_{2}X_{5}+x_{2}Y_{5}B_{12}%
}{x_{2}(Y_{5}B_{16}-X_{4})}  \nonumber \\
D_{16} &=&-\frac{%
(y_{5}Y_{6}+x_{2}Y_{5}B_{15})E_{05}+y_{5}X_{2}+F_{3}x_{2}X_{5}+x_{2}Y_{5}B_{13}%
}{x_{2}(Y_{5}B_{16}-X_{4})},  \nonumber \\
D_{17} &=&-\frac{X_{5}+Y_{5}B_{14}}{Y_{5}B_{16}-X_{4}}
\end{eqnarray}
\end{itemize}

and%
\begin{eqnarray}
B_{3}b_{1}
&=&d_{21}b_{1}B_{3}+d_{22}b_{1}B_{1}+d_{23}b_{2}D_{1}+d_{24}b_{2}D_{2}+d_{25}b_{2}D_{3}+d_{26}B_{2}d_{1}+d_{27}B_{2}d_{2}
\nonumber \\
C_{1}c_{3}
&=&D_{21}c_{3}C_{1}+D_{22}c_{1}C_{1}+D_{23}d_{1}C_{2}+D_{24}d_{2}C_{2}+D_{25}d_{3}C_{2}+D_{26}D_{1}c_{2}+D_{27}D_{2}c_{2}
\nonumber \\
&&
\end{eqnarray}%
where%
\begin{eqnarray}
d_{21} &=&-\frac{X_{4}}{X_{3}}+\frac{Y_{5}}{X_{3}}B_{16},\quad d_{22}=\frac{%
y_{5}}{x_{2}}\frac{Y_{6}}{X_{3}}-\frac{y_{5}}{x_{2}}\frac{Y_{6}}{X_{3}}%
e_{01}+\frac{Y_{5}}{X_{3}}B_{15}  \nonumber \\
d_{23} &=&F_{1}\frac{y_{5}}{x_{2}}\frac{X_{2}}{X_{3}}+F_{2}\frac{X_{5}}{X_{3}%
}-\frac{y_{5}}{x_{2}}\frac{Y_{6}}{X_{3}}e_{03}+\frac{Y_{5}}{X_{3}}B_{12} 
\nonumber \\
d_{24} &=&\frac{y_{5}}{x_{2}}\frac{X_{2}}{X_{3}}+F_{3}\frac{X_{5}}{X_{3}}-%
\frac{y_{5}}{x_{2}}\frac{Y_{6}}{X_{3}}e_{02}+\frac{Y_{5}}{X_{3}}B_{13},\quad
d_{25}=\frac{X_{5}}{X_{3}}+\frac{Y_{5}}{X_{3}}B_{14}  \nonumber \\
d_{26} &=&-f_{1}\frac{y_{5}}{x_{2}}\frac{X_{2}}{X_{3}}-\frac{y_{5}}{x_{2}}%
\frac{Y_{6}}{X_{3}}e_{04}+\frac{Y_{5}}{X_{3}}B_{11},\quad  \nonumber \\
d_{27} &=&-\frac{y_{5}}{x_{2}}\frac{X_{2}}{X_{3}}-\frac{y_{5}}{x_{2}}\frac{%
Y_{6}}{X_{3}}e_{05}
\end{eqnarray}

\begin{itemize}
\item $(F_{28},f_{28})$%
\begin{eqnarray}
D_{2}b_{2}
&=&b_{21}b_{2}D_{2}+b_{22}B_{2}d_{1}+b_{23}B_{2}d_{2}+b_{24}B_{2}d_{3}+b_{25}B_{1}b_{1}+b_{26}B_{1}b_{3}
\nonumber \\
&&+b_{27}B_{3}b_{1}+b_{28}B_{3}b_{3}  \nonumber \\
C_{2}d_{2}
&=&B_{21}d_{2}C_{2}+B_{22}D_{1}c_{2}+B_{23}D_{2}c_{2}+B_{24}D_{3}c_{2}+B_{25}C_{1}c_{1}+B_{26}C_{3}c_{1}
\nonumber \\
&&+B_{27}C_{1}c_{3}+B_{28}C_{3}c_{3}
\end{eqnarray}%
where%
\begin{eqnarray}
B_{21} &=&1+\frac{x_{5}}{x_{2}}\frac{X_{3}}{X_{2}}d_{27}+\frac{Y_{6}}{X_{2}}%
e_{05},\quad B_{22}=-f_{1}B_{12}+\frac{x_{5}}{x_{2}}\frac{X_{3}}{X_{2}}%
d_{23}+\frac{Y_{6}}{X_{2}}e_{03}  \nonumber \\
B_{23} &=&-f_{1}B_{13}+\frac{x_{5}}{x_{2}}\frac{X_{3}}{X_{2}}d_{24}+\frac{%
Y_{6}}{X_{2}}e_{02},\quad B_{24}=-f_{1}B_{14}+\frac{x_{5}}{x_{2}}\frac{X_{3}%
}{X_{2}}d_{25}  \nonumber \\
B_{25} &=&-f_{1}B_{15}+\frac{x_{5}}{x_{2}}\frac{X_{3}}{X_{2}}d_{22}+\frac{%
Y_{6}}{X_{2}}e_{01},\quad B_{26}=-f_{1}B_{16}+\frac{x_{5}}{x_{2}}\frac{X_{3}%
}{X_{2}}d_{21}  \nonumber \\
B_{27} &=&-\frac{y_{5}}{x_{2}}\frac{X_{3}}{X_{2}},\quad B_{28}=\frac{X_{6}}{%
X_{2}}
\end{eqnarray}

\item $(FF_{35},ff_{35})$%
\begin{eqnarray}
C_{3}b_{1}
&=&c_{21}b_{1}C_{1}+c_{22}b_{1}C_{3}+c_{23}B_{1}c_{3}+c_{24}B_{3}c_{3}+c_{25}b_{2}C_{2}+c_{26}d_{1}D_{1}
\nonumber \\
&&+c_{27}d_{1}D_{2}+c_{28}d_{1}D_{3}+c_{29}D_{1}d_{1}+c_{210}D_{1}d_{2}+c_{211}D_{2}d_{1}+c_{212}D_{2}d_{2}
\nonumber \\
&&+c_{213}D_{3}d_{1}+c_{214}D_{3}d_{2}  \nonumber \\
C_{1}b_{3}
&=&C_{21}b_{1}C_{1}+C_{22}b_{3}C_{1}+C_{23}B_{3}c_{1}+C_{24}B_{3}c_{3}+C_{25}b_{2}C_{2}+C_{26}d_{1}D_{1}
\nonumber \\
&&+C_{27}d_{2}D_{1}+C_{28}d_{3}D_{1}+C_{29}D_{1}d_{1}+C_{210}D_{2}d_{1}+C_{211}D_{1}d_{2}
\nonumber \\
&&+C_{213}D_{1}d_{3}+C_{214}D_{2}d_{3}
\end{eqnarray}%
where%
\begin{eqnarray}
c_{21} &=&\frac{y_{6}}{x_{3}}\frac{Y_{5}}{X_{2}}(c_{11}-1),\quad c_{22}=-%
\frac{x_{4}}{x_{3}}+\frac{y_{6}}{x_{3}}\frac{Y_{5}}{X_{2}}c_{12},\quad
c_{23}=-\frac{y_{7}}{x_{3}}+\frac{y_{6}}{x_{3}}\frac{Y_{5}}{X_{2}}c_{13} 
\nonumber \\
c_{24} &=&\frac{y_{6}}{x_{3}}\frac{X_{5}}{X_{2}}+\frac{y_{6}}{x_{3}}\frac{%
Y_{5}}{X_{2}}c_{14},\quad c_{25}=-\frac{y_{6}}{x_{3}}\frac{X_{1}}{X_{2}}+%
\frac{y_{6}}{x_{3}}\frac{Y_{5}}{X_{2}}c_{15}  \nonumber \\
c_{26} &=&F_{1}\frac{x_{4}}{x_{3}}\frac{Y_{6}}{X_{2}}-F_{2}\frac{x_{6}}{x_{3}%
}\frac{X_{3}}{X_{2}}-\frac{y_{6}}{x_{3}}\frac{Y_{7}}{X_{2}}+\frac{y_{6}}{%
x_{3}}\frac{Y_{5}}{X_{2}}c_{16}  \nonumber \\
c_{27} &=&-F_{3}\frac{x_{6}}{x_{3}}\frac{X_{3}}{X_{2}}+\frac{x_{4}}{x_{3}}%
\frac{Y_{6}}{X_{2}}+\frac{y_{6}}{x_{3}}\frac{Y_{5}}{X_{2}}c_{17},\quad
c_{28}=-\frac{x_{6}}{x_{3}}\frac{X_{3}}{X_{2}}  \nonumber \\
c_{29} &=&f_{1}\frac{y_{7}}{x_{3}}\frac{Y_{6}}{X_{2}}-f_{1}F_{1}\frac{y_{6}}{%
x_{3}}\frac{X_{4}}{X_{2}}+f_{1}F_{2}\frac{X_{6}}{X_{2}}+\frac{y_{6}}{x_{3}}%
\frac{Y_{5}}{X_{2}}c_{18}  \nonumber \\
c_{210} &=&\frac{y_{7}}{x_{3}}\frac{Y_{6}}{X_{2}}-F_{1}\frac{y_{6}}{x_{3}}%
\frac{X_{4}}{X_{2}}+F_{2}\frac{X_{6}}{X_{2}}+\frac{y_{6}}{x_{3}}\frac{Y_{5}}{%
X_{2}}c_{19},\quad  \nonumber \\
c_{211} &=&-f_{1}\frac{y_{6}}{x_{3}}\frac{X_{4}}{X_{2}}+f_{1}F_{3}\frac{X_{6}%
}{X_{2}}+\frac{y_{6}}{x_{3}}\frac{Y_{5}}{X_{2}}c_{110}  \nonumber \\
c_{212} &=&-\frac{y_{6}}{x_{3}}\frac{X_{4}}{X_{2}}+F_{3}\frac{X_{6}}{X_{2}}+%
\frac{y_{6}}{x_{3}}\frac{Y_{5}}{X_{2}}c_{111},\quad c_{213}=f_{1}\frac{X_{6}%
}{X_{2}},\quad c_{214}=\frac{X_{6}}{X_{2}}
\end{eqnarray}

\item $(FF_{23},ff_{23})$%
\begin{eqnarray}
D_{3}b_{1}
&=&a_{31}b_{1}D_{3}+a_{32}B_{1}d_{1}+a_{33}B_{1}d_{2}+a_{34}B_{3}d_{1}+a_{35}B_{3}d_{2}+a_{36}B_{2}c_{1}
\nonumber \\
&&+a_{37}B_{2}c_{3}+a_{28}b_{2}C_{1}+a_{29}b_{2}C_{3}  \nonumber \\
C_{1}d_{3}
&=&A_{31}d_{3}C_{1}+A_{32}D_{1}c_{1}+A_{33}D_{2}c_{1}+A_{34}D_{1}c_{3}+A_{35}D_{2}c_{3}+A_{36}B_{1}c_{2}
\nonumber \\
&&+A_{37}B_{3}c_{2}+A_{28}b_{1}C_{2}+A_{29}b_{3}C_{2}
\end{eqnarray}%
where%
\begin{eqnarray}
a_{31} &=&\frac{x_{2}}{x_{3}}\frac{X_{2}}{X_{3}}-\frac{x_{2}}{x_{3}}\frac{%
Y_{6}}{X_{3}}X_{15}  \nonumber \\
a_{32} &=&-f_{1}\frac{y_{7}}{x_{3}}\frac{Y_{5}}{X_{3}}+\frac{y_{5}}{x_{3}}%
\frac{Y_{7}}{X_{3}}A_{11}-(Q_{1})a_{22}-(Q_{2})a_{12}-\frac{x_{2}}{x_{3}}%
\frac{Y_{6}}{X_{3}}X_{12}  \nonumber \\
a_{33} &=&-\frac{y_{7}}{x_{3}}\frac{Y_{5}}{X_{3}}-(Q_{2})a_{13}-(Q_{1})a_{23}
\nonumber \\
a_{34} &=&f_{1}\frac{y_{6}}{x_{3}}\frac{X_{2}}{X_{3}}-(Q_{1})a_{24}-\frac{%
x_{2}}{x_{3}}\frac{Y_{6}}{X_{3}}X_{11},\quad a_{35}=\frac{y_{6}}{x_{3}}\frac{%
X_{2}}{X_{3}}-(Q_{1})a_{25}  \nonumber \\
a_{36} &=&\frac{y_{5}}{x_{3}}\frac{Y_{7}}{X_{3}}%
A_{16}-(Q_{1})a_{26}-(Q_{2})a_{14}-\frac{x_{2}}{x_{3}}\frac{Y_{6}}{X_{3}}%
X_{18}  \nonumber \\
a_{37} &=&-\frac{y_{7}}{x_{3}}\frac{X_{1}}{X_{3}}-(Q_{2})a_{15}-(Q_{1})a_{27}
\nonumber \\
a_{38} &=&\frac{y_{5}}{x_{3}}\frac{Y_{7}}{X_{3}}%
A_{14}-(Q_{2})a_{16}-(Q_{1})a_{28}-\frac{x_{2}}{x_{3}}\frac{Y_{6}}{X_{3}}%
X_{16}  \nonumber \\
a_{39} &=&\frac{y_{5}}{x_{3}}\frac{X_{1}}{X_{3}}+\frac{y_{5}}{x_{3}}\frac{%
Y_{7}}{X_{3}}A_{15}-(Q_{1})a_{29}-\frac{x_{2}}{x_{3}}\frac{Y_{6}}{X_{3}}%
X_{17}
\end{eqnarray}
\end{itemize}

$(FF_{13},ff_{13})$%
\begin{eqnarray}
D_{3}b_{2}
&=&b_{31}B_{2}d_{3}+b_{32}B_{2}d_{1}+b_{33}B_{2}d_{2}+b_{34}b_{2}D_{3}+b_{35}B_{1}b_{1}+b_{36}B_{1}b_{3}
\nonumber \\
&&+b_{37}B_{3}b_{1}+b_{38}B_{3}b_{3}  \nonumber \\
C_{2}d_{3}
&=&B_{31}D_{3}c_{2}+B_{32}D_{1}c_{2}+B_{33}D_{2}c_{2}+B_{34}d_{3}C_{2}+B_{35}C_{1}c_{1}+B_{36}C_{3}c_{1}
\nonumber \\
&&+B_{37}C_{1}c_{3}+B_{38}C_{3}c_{3}
\end{eqnarray}%
where%
\begin{eqnarray}
b_{31} &=&-\frac{y_{7}}{x_{3}}\frac{X_{1}}{X_{3}}-(Q_{2})b_{14}-(Q_{1})b_{24}
\nonumber \\
b_{32} &=&-f_{2}\frac{y_{7}}{x_{3}}\frac{X_{1}}{X_{3}}+\frac{x_{1}}{x_{3}}%
\frac{Y_{5}}{X_{3}}D_{13}+\frac{x_{1}}{x_{3}}\frac{Y_{7}}{X_{3}}%
(B_{11}+D_{13}B_{16}+E_{03}B_{15})  \nonumber \\
&&-(Q_{2})b_{12}-(Q_{1})b_{22}  \nonumber \\
b_{33} &=&-f_{3}\frac{y_{7}}{x_{3}}\frac{X_{1}}{X_{3}}+\frac{x_{1}}{x_{3}}%
\frac{Y_{5}}{X_{3}}D_{14}+\frac{x_{1}}{x_{3}}\frac{Y_{7}}{X_{3}}%
(D_{14}B_{16}+E_{02}B_{15})-(Q_{2})b_{13}-(Q_{1})b_{23}  \nonumber \\
b_{34} &=&\frac{x_{1}}{x_{3}}\frac{X_{1}}{X_{3}}+\frac{x_{1}}{x_{3}}\frac{%
Y_{7}}{X_{3}}(B_{14}+D_{17}B_{16})+\frac{x_{1}}{x_{3}}\frac{Y_{5}}{X_{3}}%
D_{17}  \nonumber \\
b_{35} &=&\frac{x_{1}}{x_{3}}\frac{Y_{5}}{X_{3}}D_{12}+\frac{x_{1}}{x_{3}}%
\frac{Y_{7}}{X_{3}}(D_{12}B_{16}+E_{01}B_{15})-(Q_{2})b_{15}-(Q_{1})b_{25} 
\nonumber \\
b_{36} &=&-\frac{y_{7}}{x_{3}}\frac{Y_{5}}{X_{3}}-(Q_{2})b_{16}-(Q_{1})b_{26}
\nonumber \\
b_{37} &=&\frac{x_{1}}{x_{3}}\frac{Y_{5}}{X_{3}}D_{11}+\frac{x_{1}}{x_{3}}%
\frac{Y_{7}}{X_{3}}D_{11}B_{16}-(Q_{1})b_{27}  \nonumber \\
b_{28} &=&-\frac{y_{6}}{x_{3}}\frac{X_{2}}{X_{3}}-(Q_{1})b_{28}
\end{eqnarray}%
Here we have used the notation%
\begin{equation}
Q_{1}=F_{3}-\frac{y_{6}}{x_{3}}\frac{Y_{6}}{X_{3}},\qquad Q_{2}=F_{2}-F_{1}%
\frac{y_{6}}{x_{3}}\frac{Y_{6}}{X_{3}}+\frac{y_{7}}{x_{3}}\frac{Y_{7}}{X_{3}}
\end{equation}

\section{The H amplitudes}

In this appendix we summarize the expressions for the $H$-functions
presented in (\ref{baba.16}) and (\ref{baba.23}) 
\begin{eqnarray}
H_{11}(u_{p},u_{q}) &=&a_{14}(u,u_{p})\left(
c_{16}(u_{p},u_{q})+c_{18}(u_{p},u_{q})\right) +a_{15}(u,u_{p})\left(
c_{26}(u_{p},u_{q})+c_{29}(u_{p},u_{q})\right)  \nonumber \\
&&+b_{12}(u,u_{p})G_{d_{1}}(u_{p},u_{q})+\omega
(u_{p},u)a_{11}(u,u_{p})a_{12}(u,u_{q})G_{d_{1}}(u,u_{p})  \nonumber \\
H_{12}(u_{p},u_{q}) &=&a_{14}(u,u_{p})\left(
c_{17}(u_{p},u_{q})+c_{110}(u_{p},u_{q})\right) +a_{15}(u,u_{p})\left(
c_{27}(u_{p},u_{q})+c_{211}(u_{p},u_{q})\right)  \nonumber \\
&&+b_{13}(u,u_{p})G_{d_{1}}(u_{p},u_{q})+\omega
(u_{p},u)a_{11}(u,u_{p})a_{12}(u,u_{q})G_{d_{2}}(u,u_{p})  \nonumber \\
H_{13}(u_{p},u_{q})
&=&a_{14}(u,u_{p})c_{19}(u_{p},u_{q})+a_{15}(u,u_{p})c_{210}(u_{p},u_{q})+b_{12}(u,u_{p})G_{d_{2}}(u_{p},u_{q})
\nonumber \\
&&+\omega (u_{p},u)a_{11}(u,u_{p})a_{13}(u,u_{q})G_{d_{1}}(u,u_{p}) 
\nonumber \\
H_{14}(u_{p},u_{q})
&=&a_{14}(u,u_{p})c_{111}(u_{p},u_{q})+a_{15}(u,u_{p})c_{212}(u_{p},u_{q})+b_{13}(u,u_{p})G_{d_{2}}(u_{p},u_{q})
\nonumber \\
&&+\omega (u_{p},u)a_{11}(u,u_{p})a_{13}(u,u_{q})G_{d_{2}}(u,u_{p})
\end{eqnarray}%
and 
\begin{eqnarray}
H_{\alpha 1}(u_{p},u_{q}) &=&a_{j6}(u,u_{p})\left(
c_{16}(u_{p},u_{q})+c_{18}(u_{p},u_{q})\right) +a_{j7}(u,u_{p})\left(
c_{26}(u_{p},u_{q})+c_{29}(u_{p},u_{q})\right)  \nonumber \\
&&+b_{j2}(u,u_{p})G_{d_{1}}(u_{p},u_{q})+\omega
(u_{p},u)a_{j1}(u,u_{p})a_{j2}(u,u_{q})G_{d_{1}}(u,u_{p})  \nonumber \\
&&+a_{j1}(u,u_{p})a_{j4}(u,u_{q})d_{13}(u_{p},u)  \nonumber \\
H_{\alpha 2}(u_{p},u_{q}) &=&a_{j6}(u,u_{p})\left(
c_{17}(u_{p},u_{q})+c_{110}(u_{p},u_{q})\right) +a_{j7}(u,u_{p})\left(
c_{27}(u_{p},u_{q})+c_{211}(u_{p},u_{q})\right)  \nonumber \\
&&+b_{j3}(u,u_{p})G_{d_{1}}(u_{p},u_{q})+\omega
(u_{p},u)a_{j1}(u,u_{p})a_{j2}(u,u_{q})G_{d_{2}}(u,u_{p})  \nonumber \\
&&+a_{j1}(u,u_{p})a_{j4}(u,u_{q})d_{14}(u_{p},u)  \nonumber \\
H_{\alpha 3}(u_{p},u_{q})
&=&a_{j6}(u,u_{p})c_{19}(u_{p},u_{q})+a_{j7}(u,u_{p})c_{210}(u_{p},u_{q})+b_{j2}(u,u_{p})G_{d_{2}}(u_{p},u_{q})
\nonumber \\
&&+\omega
(u_{p},u)a_{j1}(u,u_{p})a_{j3}(u,u_{q})G_{d_{1}}(u,u_{p})+a_{j1}(u,u_{p})a_{j5}(u,u_{q})d_{13}(u_{p},u)
\nonumber \\
H_{\alpha 4}(u_{p},u_{q})
&=&a_{j6}(u,u_{p})c_{111}(u_{p},u_{q})+a_{j7}(u,u_{p})c_{212}(u_{p},u_{q})+b_{j3}(u,u_{p})G_{d_{2}}(u_{p},u_{q})
\nonumber \\
&&+\omega
(u_{p},u)a_{j1}(u,u_{p})a_{j3}(u,u_{q})G_{d_{2}}(u,u_{p})+a_{j1}(u,u_{p})a_{j5}(u,u_{q})d_{14}(u_{p},u)
\end{eqnarray}%
for $\alpha =2,3.$

These function satisfy the properties%
\begin{eqnarray}
H_{\alpha 1}(u_{p},u_{q}) &=&\omega (u_{p},u_{q})H_{\alpha 1}(u_{q},u_{p}) 
\nonumber \\
H_{\alpha 2}(u_{p},u_{q}) &=&\omega (u_{p},u_{q})H_{\alpha 3}(u_{q},u_{p}) 
\nonumber \\
H_{\alpha 3}(u_{p},u_{q}) &=&\omega (u_{p},u_{q})H_{\alpha 2}(u_{q},u_{p}) 
\nonumber \\
H_{\alpha 4}(u_{p},u_{q}) &=&\omega (u_{p},u_{q})H_{\alpha 4}(u_{q},u_{p})
\end{eqnarray}%
and in the quasi-classical limit of ${\cal F}_{lj}^{(n-2)}$ (\ref{off.5})
their expansion have to be considered up to $\eta ^{3}$ order: 
\begin{eqnarray}
H_{\alpha j}(u_{p},u_{q}) &=&h_{\alpha j}^{(0)}(u_{p},u_{q})+h_{\alpha
j}^{(1)}(u_{p},u_{q})\eta +\frac{1}{2}h_{\alpha j}^{(2)}(u_{p},u_{q})\eta
^{2}+\frac{1}{6}h_{\alpha j}^{(3)}(u_{p},u_{q})\eta ^{3}+{\rm o}(\eta ^{4}) 
\nonumber \\
\alpha &=&1,2,3,\quad j=1,2,3,4
\end{eqnarray}%
where $h_{\alpha j}^{(0)}(u_{p},u_{q})=h_{\alpha j}^{(1)}(u_{p},u_{q})=0.$

\end{document}